\documentclass[12pt,a4paper]{article}%
\usepackage{graphicx}
\usepackage{epsfig}
\usepackage{bm}
\usepackage{amsmath}
\usepackage{amsfonts}
\usepackage{amssymb}
\usepackage{floatflt}%
\begin{document}

\title{Adler function and hadronic contribution to the muon $g-2$ in a nonlocal
chiral quark model}
\author{Alexander E. Dorokhov\\Joint Institute for Nuclear Research, Bogoliubov Laboratory of Theoretical Physics,\\Dubna, Russia}
\maketitle

{\abstract The behavior of the vector Adler function at spacelike momenta is
studied in the framework of a covariant chiral quark model with instanton-like
quark-quark interaction. This function describes the transition between the
high energy asymptotically free region of almost massless current quarks to
the low energy hadronized regime with massive constituent quarks. The model
reproduces the Adler function and $V-A$ correlator extracted from the ALEPH
and OPAL data on hadronic $\tau$ lepton decays, transformed into the Euclidean
domain via dispersion relations. The leading order contribution from hadronic
part of the photon vacuum polarization to the anomalous magnetic moment of the
muon, $a_{\mu}^{\mathrm{hvp}\left(  1\right) } $, is estimated.}

\section{Introduction}

The transition from perturbative regime of QCD to nonperturbative one has yet
remained under discussion. At high momenta the fundamental degrees of freedom
are almost massless asymptotically free quarks. At low momenta the
nonperturbative regime is adequately described in terms of constituent quarks
with masses dynamically generated by spontaneous breaking of chiral symmetry.
The instanton model of QCD vacuum \cite{ShSh} provides the mechanism of
dynamical quark dressing in the background of instanton vacuum and leads to
generation of the momentum dependent quark mass that interpolates these two
extremes. Still it is not clear how an intuitive picture of this transition
may be tested at the level of observables. In this paper we demonstrate that
the Adler function depending on spacelike momenta may serve as the appropriate
quantity. This function defined as the logarithmic derivative of the
current-current correlator can be extracted from the experimental data of
ALEPH \cite{ALEPH2} and OPAL \cite{OPAL} collaborations on inclusive hadronic
$\tau$ decays. From theoretical point of view it is well known that in
high-energy asymptotically free limit the Adler function calculated for
massless quarks is a nonzero constant. From the other side in the constituent
quark model (suitably regularized) this function is zero at zero virtuality.
Thus the transition of the Adler function from its constant asymptotic
behaviour to zero is very indicative concerning the nontrivial QCD dynamics at
intermediate momenta. In this paper we intend to show that the instanton-like
nonlocal chiral quark model (N$\chi$QM) describes this transition correctly.
In particular, we analyze the correlator of vector currents and corresponding
Adler function in the framework of N$\chi$QM that allows us to draw a precise
and unambiguous comparison of the experimental data with the model
calculations. The use in the calculations of a covariant nonlocal low-energy
quark model based on the self-consistent approach to the dynamics of quarks
has many attractive features as it preserves the gauge invariance, is
consistent with the low-energy theorems, as well as takes into account the
large-distance dynamics controlled by the bound states. As an application we
estimate the leading order hadronic vacuum polarization contribution to the
muon anomalous magnetic moment which is expressed as an convolution integral
over spacelike momenta of the Adler function and confront it with the recent
results of the measurements by the Muon $\left(  g-2\right)  $ collaboration
\cite{g-2Coll}.

The paper is organized as follows. In Sect. 2, we briefly recall the
definition of the Adler function and the way how to extract it from the
experimental data. Then in Sect. 3 we remind the definition of the leading
order contribution of hadronic vacuum polarization to the muon anomalous
magnetic moment and present its phenomenological estimates. In Sect. 4 and 5
we outline the gauged nonlocal chiral quark model extended by inclusion of the
vector and axial-vector mesons and derive the expressions for the Adler
function within the model considered. Then, after fixing the model parameters
in Sect. 6, we confront the model results with available experimental data on
the Adler function and the $V-A$ correlator in Sects. 7 and 8,
correspondingly. Sect. 9 contains our conclusions. In Appendices we give
necessary information about the nonlocal vertices of quark interaction with
external currents, phenomenology of vector mesons and the structure of
non-chiral corrections to low energy observables.

\section{\noindent The Adler function}

In the chiral limit, where the masses of $u$, $d$, $s$ light quarks are set to
zero, the vector ($V$) and non-singlet axial-vector ($A$) current-current
correlation functions in the momentum space (with $-q^{2}\equiv Q^{2}\geq0$)
are defined as
\begin{align}
\Pi_{\mu\nu}^{J,ab}(q)  &  =i\int d^{4}x~e^{iqx}\Pi_{\mu\nu}^{J,ab}%
(x)=\,\left(  q_{\mu}q_{\nu}-g_{\mu\nu}q^{2}\right)  \Pi_{J}(Q^{2})\delta
^{ab},\label{PA}\\
\qquad\Pi_{\mu\nu}^{J,ab}(x)  &  =\langle0\left\vert T\left\{  J_{\mu}%
^{a}(x)J_{\nu}^{b}(0)^{\dagger}\right\}  \right\vert 0\rangle,\nonumber
\end{align}
where the QCD $V$ and $A$ currents are%
\begin{equation}
J_{\mu}^{a}=\overline{q}\gamma_{\mu}\frac{\lambda^{a}}{\sqrt{2}}q,\qquad
J_{\mu}^{5a}=\overline{q}\gamma_{\mu}\gamma_{5}\frac{\lambda^{a}}{\sqrt{2}}q,
\label{JAV}%
\end{equation}
and $\lambda^{a}$ are Gell-Mann matrices $\left(  \mathrm{tr}\lambda
^{a}\lambda^{b}=2\delta^{ab}\right)  $. The momentum-space two-point
correlation functions obey (suitably subtracted) dispersion relations,
\begin{equation}
\Pi_{J}(Q^{2})=\int_{0}^{\infty}\frac{ds}{s+Q^{2}}\frac{1}{\pi}\mathrm{Im}%
\Pi_{J}(s), \label{Peuclid}%
\end{equation}
where the imaginary parts of the correlators determine the spectral functions%
\begin{equation}
\rho_{J}(s)=4\pi\mathrm{\operatorname{Im}}\Pi_{J}(s+i0).\nonumber
\end{equation}
Instead of the correlation function it is more convenient to work with the
Adler function defined as
\begin{equation}
D_{J}(Q^{2}){=-Q}^{2}\frac{d\Pi_{J}(Q^{2})}{dQ^{2}}{\,=}\frac{1}{4\pi^{2}}%
\int_{0}^{\infty}{dt}\frac{Q^{2}}{(t+Q^{2})^{2}}{\,\rho_{J}(t)\,.}
\label{Adler}%
\end{equation}

Recently, the isovector $V$ and $A$ spectral functions have been determined
separately with high precision by the ALEPH \cite{ALEPH2} and OPAL \cite{OPAL}
collaborations from the inclusive hadronic $\tau$-lepton decays ($\tau
\rightarrow\nu_{\tau}+$ hadrons) in the interval of invariant masses up to the
$\tau$ mass, $0\leq s\leq m_{\tau}^{2}$. It is important to note that the
experimental separation of the $V$ and $A$ spectral functions allows us to
test accurately the saturation of the chiral sum rules of Weinberg-type in the
measured interval. On the other hand, at large $s$ the correlators can be
confronted with perturbative QCD (pQCD) thanks to sufficiently large value of
the $\tau$ mass. \begin{figure}[h]
\hspace*{1cm} \begin{minipage}{7cm}
\vspace*{0.5cm} \epsfxsize=6cm \epsfysize=5cm \centerline{\epsfbox
{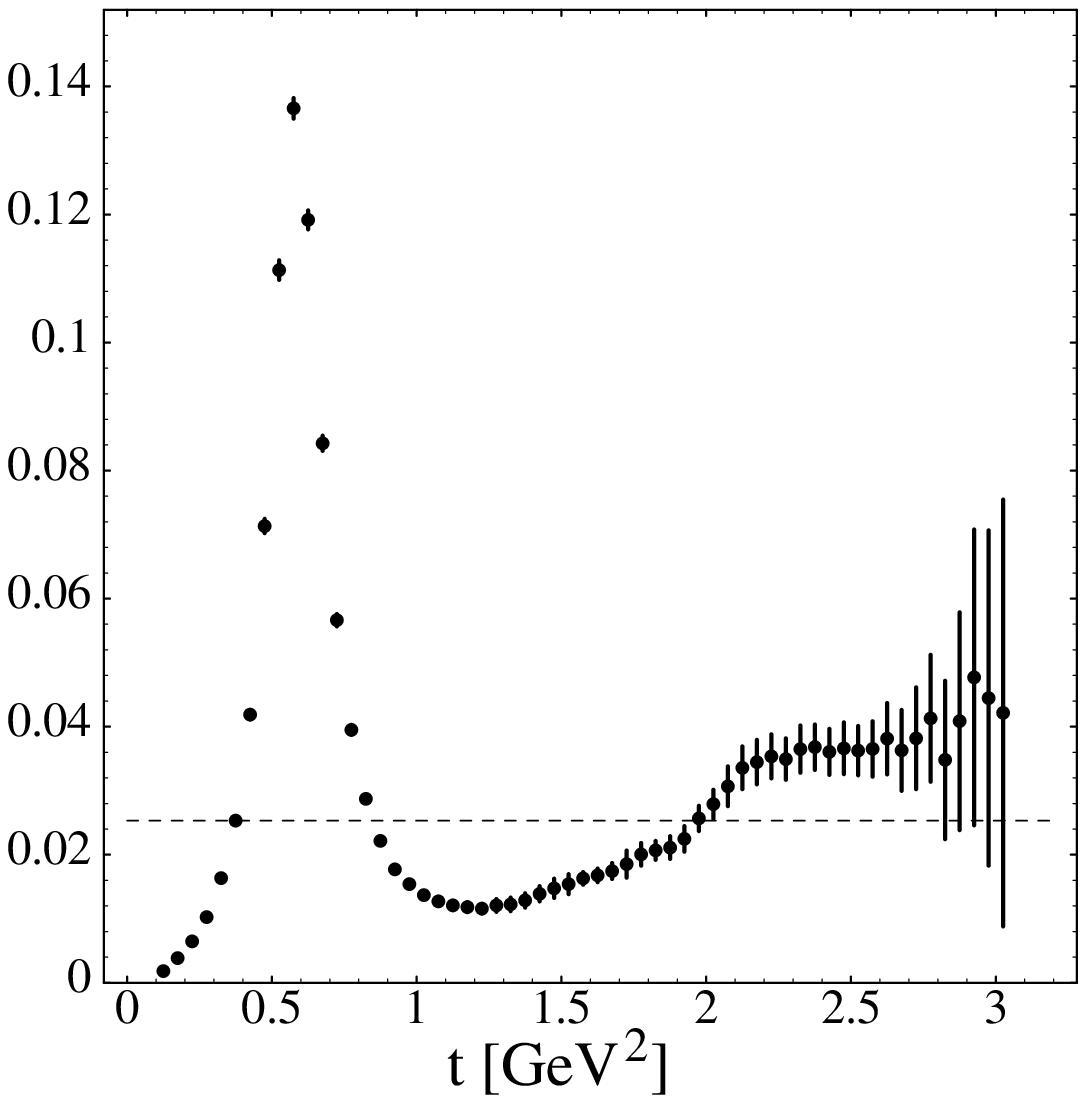}}
\caption[dummy0]{The isovector vector  spectral function from hadronic
$\tau$- decays [2]. The dashed line is the asymptotic freedom prediction,
$1/(4\pi^{2}).$ \label{FAleph} }
\end{minipage}\hspace*{0.5cm} \begin{minipage}{7.2cm}
\vspace*{0.5cm} \epsfxsize=6cm \epsfysize=5cm \centerline{\epsfbox
{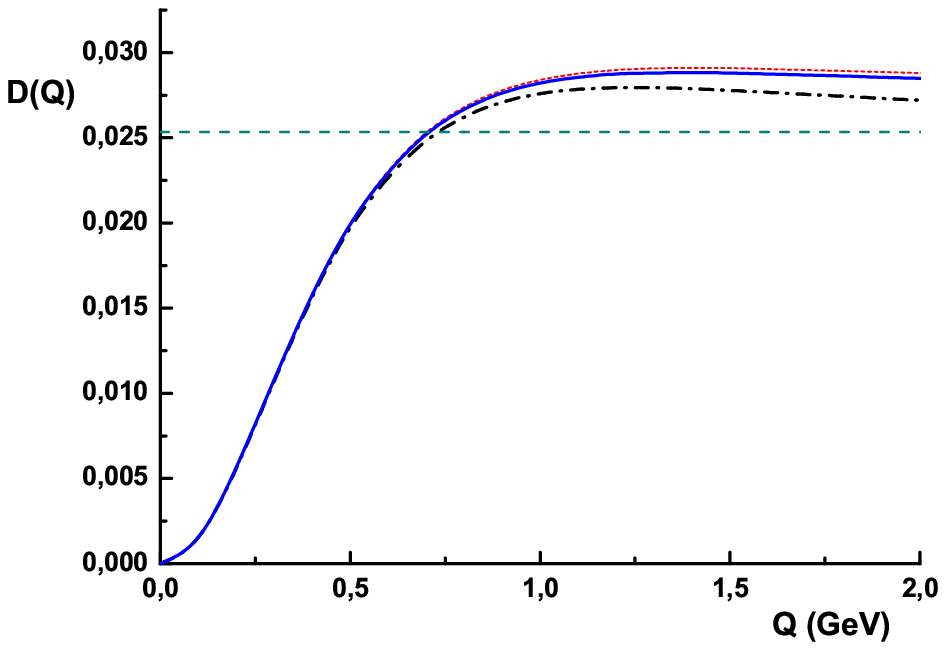}}
\caption[dummy0]{The vector Adler function constructed with use of
LO (dot-dashed), NLO
(dotted) and NNLO (full line) pQCD asymptotics. The dashed line is asymptotic
freedom prediction, $1/4\pi^{2}$. \label{AdlerALEPH} }
\end{minipage}\end{figure}

The vector spectral function and the corresponding Adler function determined
from the ALEPH data (see below) are shown in Figs. \ref{FAleph} and
\ref{AdlerALEPH}.
%\begin{figure}[h]
%\includegraphics[height=7.5cm]{Figppr3.ps}\caption{The isovector hadronic
%spectral function from hadronic $\tau$- decays. The dashed line is the
%asymptotic freedom prediction $1/(4\pi^{2}).$}%
%\label{FAleph}%
%\end{figure}
%\begin{figure}[h]
%\includegraphics[height=7.5cm]{AdlerALEPH.eps}\caption{LO (dot-dashed), NLO
%(dotted) and NNLO (full line) pQCD asymptotics. The dashed line is asymptotic
%freedom prediction $1/4\pi^{2}$.}%
%\label{AdlerALEPH}%
%\end{figure}
The behaviour of the correlators at low and high momenta is constrained by
QCD. In the regime of large momenta the Adler function is dominated by pQCD
contribution supplemented by small power corrections
\begin{equation}
D_{V}(Q^{2}\rightarrow\infty)=D_{V}^{\mathrm{pQCD}}(Q^{2})-\frac{\alpha_{s}%
}{4\pi^{3}}\frac{\lambda^{2}}{Q^{2}}+\frac{1}{6}\frac{\alpha_{s}}{\pi}%
\frac{\left\langle \left(  G_{\mu\nu}^{a}\right)  ^{2}\right\rangle }{Q^{4}%
}+\frac{O_{D}^{6}}{Q^{6}}+\mathcal{O}(\frac{1}{Q^{8}}), \label{Dope}%
\end{equation}
where the pQCD contribution with three-loop accuracy is given in the chiral
limit in $\overline{\mathrm{MS}}$ renormalization scheme by
\cite{3Loop,Chet3}
\begin{equation}
D_{V}^{\mathrm{pQCD}}(Q^{2};\mu^{2})={\frac{1}{4\pi^{2}}}\left\{
1+\frac{\alpha_{s}\left(  \mu^{2}\right)  }{\pi}+\left[  F_{2}-\beta_{0}%
\ln{\frac{Q^{2}}{\mu^{2}}}\right]  \left(  \frac{\alpha_{s}(\mu^{2})}{\pi
}\right)  ^{2}\right.  + \label{DpQCD}%
\end{equation}%
\[
{+}\left.  \left[  F_{3}-\left(  {2}F_{2}\beta_{0}{+\beta_{1}}\right)
{\ln{\frac{Q^{2}}{\mu^{2}}}+\beta_{0}^{2}}\left(  \frac{\pi^{2}}{3}{+\ln}%
^{2}{{\frac{Q^{2}}{\mu^{2}}}}\right)  \right]  \left(  \frac{\alpha_{s}%
(\mu^{2})}{\pi}\right)  ^{3}{+}\mathcal{O}{(\alpha}_{s}^{4}{)}\right\}
\]
where
\[
\beta_{0}{=}\frac{1}{4}\left(  11{-}\frac{2}{3}n_{f}\right)  {\,,\qquad}%
\beta_{1}{=}\frac{1}{8}\left(  51{-}\frac{19}{3}{n}_{f}\right)  {\,,}%
\]%
\[
F_{2}{=1.98571-0.115295\,n}_{f}{\,,\qquad}F_{3}{=-6.63694-1.20013\,n}%
_{f}{-0.00518\,n}_{f}^{2}{\,,}%
\]
with $\alpha_{s}(Q^{2})$ being the solution of the equation
\begin{equation}
\frac{\pi}{\beta_{0}\alpha_{s}(Q^{2})}-\frac{\beta_{1}}{\beta_{0}^{2}}%
\ln\left[  \frac{\pi}{\beta_{0}\alpha_{s}(Q^{2})}+\frac{\beta_{1}}{\beta
_{0}^{2}}\right]  =\ln\frac{Q^{2}}{\Lambda^{2}}\,. \label{AlphaRG}%
\end{equation}
In (\ref{Dope}) along with standard power corrections due to the gluon and
quark condensates we include the unconventional term suppressed as,
$\sim1/Q^{2}$. Its appearance was augmented in \cite{ChetNar} and also found
in the N$\chi$QM \cite{DoBr03}.

In the low-$Q^{2}$ limit it is only rigorously known from the theory that
\begin{equation}
{D}_{V}{(Q}^{{2}}{\rightarrow0)=Q}^{{2}}{D}_{V}^{\prime}{(0)+}\mathcal{O}%
{(Q}^{4}{).} \label{D(0)}%
\end{equation}
It is clear (see also Fig. \ref{AdlerALEPH}) that the Adler function is very
sensitive to transition between asymptotically free (almost massless current
quarks) region described by (\ref{Dope}), (\ref{DpQCD}) to the hadronic regime
with almost constant constituent quarks where one has (\ref{D(0)}).

\begin{figure}[h]
\hspace*{1cm} \begin{minipage}{7cm}
\vspace*{0.5cm} \epsfxsize=6cm \epsfysize=5cm \centerline{\epsfbox{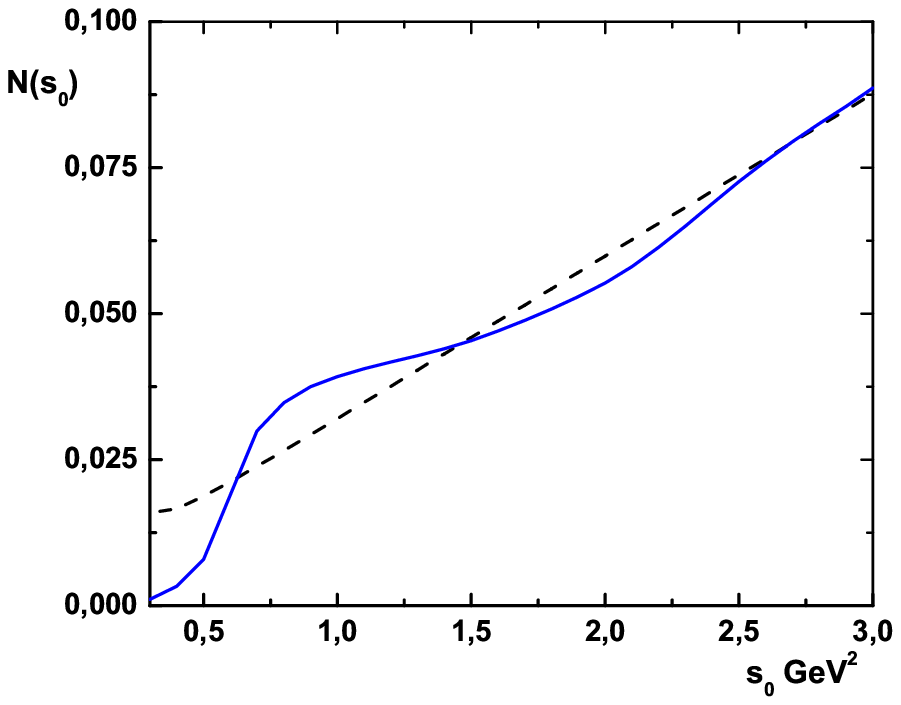}}
\caption[dummy0]{
The integral, Eq.
(\ref{Matching}), versus the upper integration limit, $s_{0}$, for the $V$
spectral density. The integral of the experimental data corresponds to solid line
and the pQCD prediction (\ref{MatchQCD}) is given by the dashed line.
\label{NormV} }
\end{minipage}\hspace*{0.5cm} \begin{minipage}{7cm}
\vspace*{0.5cm} \epsfxsize=6cm \epsfysize=5cm \centerline{\epsfbox
{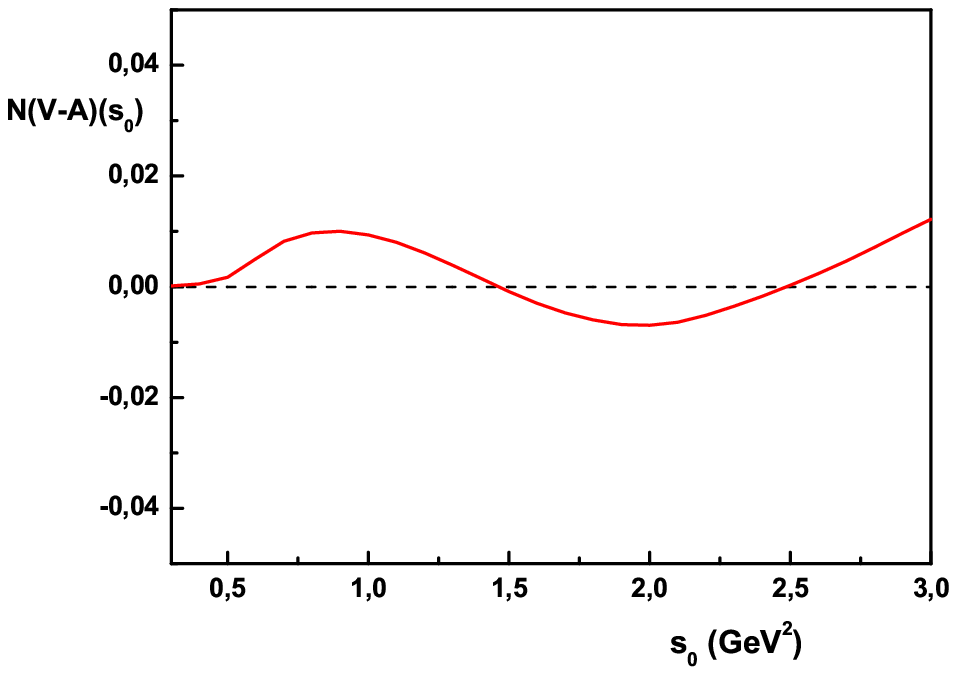}}
\caption[dummy0]{
The integral, Eq.
(\ref{Matching}), versus the upper integration limit, $s_{0}$, for the $V-A$
spectral density (second Weinberg sum rule). The pQCD prediction is given by
dashed line, and the experimental function is given by solid line.  \label
{NormV-A} }
\end{minipage}\end{figure}

To extract the Adler function from experimental data supplemented by QCD
asymptotics (\ref{Dope}), (\ref{DpQCD}) we take following \cite{PPdR98} an
ansatz for the hadronic spectral functions in the form
\begin{equation}
\rho_{J}\left(  t\right)  =\rho_{\mathrm{J}}^{\mathrm{ALEPH}}\left(  t\right)
\theta(s_{0}-t)+\rho_{J}^{\mathrm{pQCD}}\left(  t\right)  \theta(t-s_{0})\,\,,
\label{SpecDens}%
\end{equation}
where
\begin{equation}
\frac{1}{4\pi^{2}}\rho_{V}^{\mathrm{pQCD}}\left(  t\right)  =D_{V}%
^{\mathrm{pQCD}}\left(  t\right)  -\frac{121\pi^{2}}{48}\left(  \frac
{\alpha_{s}(t)}{\pi}\right)  ^{3}, \label{SpecDensQCD}%
\end{equation}
and find the value of continuum threshold $s_{0}$ from the global duality
interval condition:
\begin{equation}
\int_{0}^{s_{0}}dt\,\rho_{\mathrm{J}}^{\mathrm{ALEPH}}(t)=\int_{0}^{s_{0}%
}dt\rho_{J}^{\mathrm{pQCD}}\,(t)\,. \label{Matching}%
\end{equation}
Using the experimental input corresponding to the $\tau$--decay data and the
pQCD expressions~
\begin{align}
\frac{1}{4\pi^{2}}\int_{0}^{s_{0}}dt\,\rho_{V}^{\mathrm{pQCD}}(t)  &
=\frac{N_{c}}{12\pi^{2}}s_{0}\left\{  1+\frac{\alpha_{\mathrm{s}}(s_{0})}{\pi
}+\left[  F_{2}+\beta_{0}\right]  \left(  \frac{\alpha_{\mathrm{s}}(s_{0}%
)}{\pi}\right)  ^{2}+\right. \label{MatchQCD}\\
&  \left.  +\left[  F_{3}+\left(  2F_{2}\beta_{0}+\beta_{1}\right)
+2\beta_{0}^{2}\right]  \left(  \frac{\alpha_{\mathrm{s}}(s_{0})}{\pi}\right)
^{3}\right\}  ,\nonumber\\
\rho_{V-A}^{\mathrm{pQCD}}(t)\,  &  =0, \label{IIWSR}%
\end{align}
one finds (see Figs. \ref{NormV} and \ref{NormV-A}) that matching between the
experimental data and theoretical predictions occurs approximately at scale
$s_{0}\approx2.5\mathrm{GeV}^{2}$. Note that the condition (\ref{IIWSR}) in
the $V-A$ channel corresponds to matching the second Weinberg chiral sum rule.

The vector Adler function (\ref{Adler}) obtained from matching the low momenta
experimental data and high momenta pQCD asymptotics by using the spectral
density (\ref{SpecDens}) is shown in Fig. \ref{AdlerALEPH}, where we use the
pQCD asymptotics (\ref{SpecDensQCD}) of the massless vector spectral function
to four loops with $\Lambda_{\overline{\mathrm{MS}}}^{n_{f}=3}=372$ MeV and
choose the matching parameter as $s_{0}=2.5$ GeV$^{-1}.$ Admittedly, in the
Euclidean presentation of the data the detailed resonance structure
corresponding to the $\rho$ and $a_{1}$ mesons seen in the Minkowski region
(Fig. \ref{FAleph}) is smoothed out, hence the verification of the theory is
not as stringent as would be directly in the Minkowski space.

\section{Leading order hadronic vacuum polarization contributions.}

The anomalous magnetic moment of the muon is known to an unprecedented
accuracy of order of $1$ ppm. The latest result from the measurements of the
Muon $(g-2)$ collaboration at Brookhaven is \cite{g-2Coll}
\begin{equation}
a_{\mu}\equiv{\frac{1}{2}}(g_{\mu}-2)=11\ 659\ 208\left(  6\right)
\cdot10^{-10}. \label{AMMg-2}%
\end{equation}

Using $e^{+}e^{-}$ annihilation data and data from hadronic $\tau$ decays the
standard model predictions are \cite{DEHZh03,J}
\begin{equation}
a_{\mu}^{SM}=\left\{
\begin{array}
[l]{l}%
11\ 659\ 181\left(  8\right)  \cdot10^{-10}\qquad e^{+}e^{-},\\
11\ 659\ 196\left(  7\right)  \cdot10^{-10}\qquad\tau.
\end{array}
\right.  \label{AMMee}%
\end{equation}
%\label{AMMtau}

The difference between the experimental determination of $a_{\mu}$ and the
standard model using the $e^{+}e^{-}$ or $\tau$ data for the calculation of
the hadronic vacuum polarization is $2.7\sigma$ and $1.4\sigma$, respectively.

The standard model prediction for $a_{\mu}$ consists of quantum
electrodynamics, weak and hadronic contributions. The QED and weak
contributions to $a_{\mu}$ have been calculated with great accuracy
\cite{KinNio04}%
\begin{equation}
a_{\mu}^{\mathrm{QED}}=11~658~471.935(0.203)\cdot10^{-10} \label{AMMqed}%
\end{equation}
and \cite{CzMV03}
\begin{equation}
a_{\mu}^{\mathrm{EW}}=15.4(0.3)\cdot10^{-10}. \label{AMMweak}%
\end{equation}

The uncertainties of the standard model values in (\ref{AMMee}) are dominated
by the uncertainties of the hadronic photon vacuum polarization. Thus, to
confront usefully theory with the experiment requires a better determination
of the hadronic contributions. In the last decade, a substantial improvement
in the accuracy of the contribution from the hadronic vacuum polarization was
reached. It uses, essentially, precise determination of the low energy tail of
the total $e^{+}e^{-}\rightarrow$ hadrons and $\tau$ lepton decays
cross-sections. The contributions of hadronic vacuum polarization at order
$\alpha^{2}$ quoted in the most recent articles on the subject are given in
the Table 1.\\[0.1cm]

Table 1.\\[0.1cm]%

\begin{tabular}
[c]{|c|c|c|c|c|c|}\hline
& $e^{+}e^{-}\cite{DEHZh03}$ & $\tau\cite{DEHZh03}$ & $e^{+}e^{-}\cite{J}$ &
$e^{+}e^{-}\cite{TrocYnd}$ & $\tau\cite{TrocYnd}$\\\hline
$a_{\mu}^{\mathrm{hvp}~\left(  1\right)  }\cdot10^{10}$ & $696.3\pm9.8$ &
$711.0\pm8.6$ & $694.8\pm8.6$ & $693.5\pm9.0$ & $701.8\pm8.9$\\\hline
\end{tabular}
\\[0.3cm]

The higher-order contributions at $O(\alpha^{3})$ level to $a_{\mu
}^{\mathrm{hvp}~\left(  2\right)  }$ was estimated in \cite{Kra97}
\begin{equation}
a_{\mu}^{\mathrm{hvp}~\left(  2\right)  }=-10.1(0.6)\cdot10^{-10} \label{AMM2}%
\end{equation}
by using analytical kernel functions and experimental data on the $e^{+}%
e^{-}\rightarrow$ hadrons cross-section. In addition, there exists the
$O(\alpha^{3})$ contribution to $a_{\mu}$ from the hadronic light-by-light
scattering diagram, $a_{\mu}^{\mathrm{h.~L\times L}}$, that cannot be
expressed as a convolution of experimentally accessible observables and need
to be estimated from theory. The recent estimate of the hadronic
light-by-light scattering contribution reads \cite{knecht02}
\begin{equation}
a_{\mu}^{\mathrm{h.~L\times L}}=8(4)\cdot10^{-10}\ . \label{AMM_LL}%
\end{equation}
The latest estimate of this term is given in \cite{MelnVain03}, where strong
constraints on the light-by-light scattering amplitude from the short-distance
QCD have been imposed, with the result $a_{\mu}^{\mathrm{h.~L\times L}%
}=13.6(2.5)\cdot10^{-10}$.

Phenomenological estimate of the total hadronic contributions to $a_{\mu
}^{\mathrm{hvp}~\left(  2\right)  }$ has to be compared with the value deduced
from the $g-2$ experiment (\ref{AMMg-2}) and known electroweak and QED
corrections
\begin{equation}
a_{\mu}^{\mathrm{hvp}}=720.7\left(  6.0\right)  \cdot10^{-10}.
\label{aMMH1EXP}%
\end{equation}
The agreement between the standard model prediction and the present
experimental value is rather good. There is certain mismatch between the
experimental and theoretical predictions for $a_{\mu}$, but in view of the
inconsistencies between the evaluations based on $e^{+}e^{-}$ and $\tau$ data
the conclusion about discrepancy of the experiment and standard model is
certainly premature.

\begin{floatingfigure}[l]{0.4\textwidth}
\begin{center}
\resizebox{0.3\textwidth}{0.3\textwidth}{\includegraphics{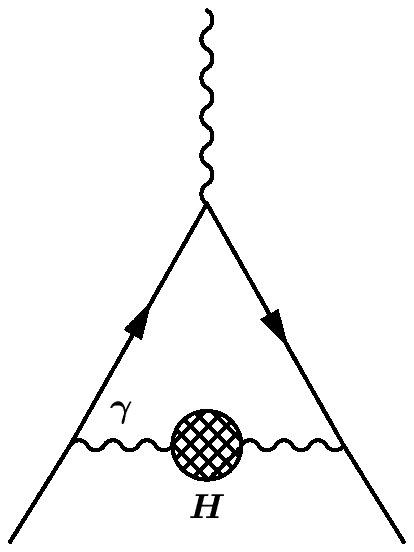}}
\caption{The contribution of the hadronic photon vacuum polarization to anomalous magnetic moment.}
\label{AMMloFig}\end{center}
\end{floatingfigure}

%\begin{figure}[h]
%\begin{minipage}{7cm}
%%\hspace*{1.cm}
%\vspace*{0.5cm}
%%\epsfxsize=5cm \epsfysize=5cm
%\centerline{\epsfbox{Fig-g2a.eps}}
%\caption[dummy0]
%{The contribution of the
%hadronic photon vacuum polarization to anomalous magnetic moment.
%\label{AMMloFig}}
%\end{minipage}\end{figure}

In this work we analyze the contribution of hadronic photon vacuum
polarization at order $\alpha^{2}$ to $a_{\mu}^{\mathrm{hvp}}$ (Fig.
\ref{AMMloFig}) from the point of view of the nonlocal chiral quark model of
low energy QCD and show that, within this framework, it might be possible
realistically to determine this value to a sufficiently safe accuracy. We want
to discuss how well this model, which has been developed in refs.
\cite{ADoLT00} and \cite{DoBr03}, does in calculating $a_{\mu}^{\mathrm{hvp}%
~\left(  1\right)  }$. This quantity is usually expressed in the form of a
spectral representation $(e^{2}=4\pi\alpha)$
\begin{equation}
a_{\mu}^{\mathrm{hvp}\left(  1\right)  }=\left(  \frac{\alpha}{\pi}\right)
^{2}\int_{0}^{\infty}dt\frac{1}{t}K(t)\rho_{\mathrm{V}}^{(\mathrm{H})}\left(
t\right)  \ , \label{Amm_rho}%
\end{equation}
which is a convolution of the hadronic spectral function $\rho_{V}%
^{\mathrm{(H)}}\left(  t\right)  $, related to the total $e^{+}e^{-}%
\rightarrow\gamma^{\ast}\rightarrow$ hadrons cross-section $\sigma(t)$
by$(m_{e}\rightarrow0)$
\begin{equation}
\sigma(t)=4\pi\alpha^{2}\frac{1}{t}\rho_{\mathrm{V}}^{(\mathrm{H})}\left(
t\right)  , \label{Sigma_rho}%
\end{equation}
with the QED function
\begin{equation}
K(t)=\int_{0}^{1}dx{\frac{x^{2}(1-x)}{x^{2}+(1-x)t/m_{\mu}^{2}}}\ ,
\label{Kfac}%
\end{equation}
which is sharply peaked at low $t$ and decreases monotonically with increasing
$t$.

For our purposes, it is convenient to express $a_{\mu}^{\mathrm{hvp}\left(
1\right)  }$ using the integral representation \cite{6} in terms of the Adler
function%
\begin{equation}
a_{\mu}^{\mathrm{hvp}\left(  1\right)  }=\frac{4}{3}\alpha^{2}\int_{0}%
^{1}dx\frac{\left(  1-x\right)  \left(  2-x\right)  }{x}D_{V}\left(
\frac{x^{2}}{1-x}m_{\mu}^{2}\right)  , \label{aAd}%
\end{equation}
where the charge factor $\sum Q_{i}^{2}=2/3$, $i=u,d,s,$ is taken into
account. By using the Adler function determined from experiment (\ref{Adler})
and (\ref{SpecDens}) one gets the estimate
\begin{equation}
a_{\mu}^{\mathrm{hvp}\left(  1\right)  }=7.22\cdot10^{-8} \label{ammALEPH}%
\end{equation}
which is in a reasonable agreement with the precise phenomenological numbers
quoted in Table 1. In the following we determine the Adler function and
$a_{\mu}^{\mathrm{hvp}\left(  1\right)  }$ from the effective quark model
describing the dynamics of low and intermediate energy QCD.

\section{The extended nonlocal chiral quark model}

The bulk of the integral in (\ref{aAd}) is governed by the low energy
behaviour of the Adler function $D_{V}(Q^{2})$. The typical momentum of the
virtual photon in Fig. \ref{AMMloFig} is $Q^{2}\sim m_{\mu}^{2}$. These are
momenta values much smaller than the characteristic scales of the spontaneous
chiral symmetry breaking or confinement $(\simeq1GeV)$. Therefore, the
appropriate way to look at this problem is within the framework of the low
energy effective field model of QCD. In the low momenta domain the effect of
the nonperturbative structure of QCD vacuum become dominant. Since invention
of the QCD sum rule method based on the use of the standard operator product
expansion (OPE) it is common to parameterize the nonperturbative properties of
the QCD vacuum by using infinite towers of the vacuum average values of the
quark-gluon operators. From this point of view the nonlocal properties of the
QCD vacuum result from the partial resummation of the infinite series of power
corrections, related to vacuum averages of quark-gluon operators with growing
dimension, and may be conventionally described in terms of the nonlocal vacuum
condensates \cite{MikhRad92,DEM97}. This reconstruction leads effectively to
nonlocal modifications of the propagators and effective vertices of the quark
and gluon fields. The adequate model describing this general picture is the
instanton liquid model of QCD vacuum describing nonperturbative nonlocal
interactions in terms of the effective action \cite{ShSh}. Spontaneous
breaking the chiral symmetry and dynamical generation of a momentum-dependent
quark mass are naturally explained within the instanton liquid model. The $V$
and $A$ current-current correlators have been calculated in \cite{DoBr03} in
the framework of the effective chiral model with instanton-like nonlocal
quark-quark interactions \cite{ADoLT00} (N$\chi$QM). In the present work we
extend that analysis by inclusion into consideration of the vector and
axial-vector mesons generated from resummation of quark loops.
%The nonlocal structure of the model is motivated by
%fundamental properties of the QCD vacuum induced by the nonperturbative gauge fields, instantons,
%which spontaneously break the chiral symmetry and generate dynamically a
%momentum-dependent quark mass.

Nonlocal effective models have an important feature which makes them
advantageous over the local models, such as the well known Nambu--Jona-Lasinio
model (NJL). At high virtualities the quark propagator and the vertex
functions of the quark coupled to external fields reduce to the free quark
propagator and to local, point-like couplings. This property allows us to
straightforwardly reproduce the leading (asymptotically free) terms of the
OPE. For instance, the second Weinberg sum rule is reproduced in the model
\cite{DoBr03,Bron99}, which has not been the case of the local approaches. In
addition, the intrinsic nonlocalities, inherent to the model, generate
unconventional power and exponential corrections which have the same character
as found in \cite{ChetNar}. The nonlocal effective model was successively
applied to the description of the data from the CLEO collaboration on the pion
transition form factor in the interval of the space-like momentum transfer
squared up to 8 GeV$^{2}$ \cite{AD02}. There are several further advantages in
using the nonlocal models compared to the local approaches, in particular, the
model is made consistent with the gauge invariance. As we shell see below the
N$\chi$QM correctly reproduces leading large $Q^{2}$ behaviour of the Adler
function, while the local constituent quark model fails to describe data
starting from rather low $Q^{2}$.

We start with the nonlocal chirally invariant action which describes the
interaction of soft quark fields. The gluon fields have been integrated out.
The corresponding gauge-invariant action for quarks interacting through
nonperturbative exchanges can be expressed as \cite{ADoLT00}
\begin{align}
S  &  =\int d^{4}x\ \overline{q}(x)\gamma^{\mu}\left[  i\partial_{\mu}-V_{\mu
}\left(  x\right)  -\gamma_{5}A_{\mu}\left(  x\right)  \right]
q(x)+\nonumber\\
&  +\frac{1}{2}\sum_{i=P,V,A}G_{i}\int d^{4}X\int\prod_{n=1}^{4}d^{4}
x_{n}\ f_{i}(x_{n})\left[  \overline{Q}(X-x_{1},X)\Gamma_{i}Q(X,X+x_{3}
)\overline{Q}(X-x_{2},X)\Gamma_{i}Q(X,X+x_{4})\right]  , \label{Lint}%
\end{align}
where in the extended version of the model the spin-flavor structure of the
interaction is given by matrix products
\begin{equation}
G_{i}\left(  \Gamma_{i}\otimes\Gamma_{i}\right)  :\qquad G_{P}\left(
1\otimes1+i\gamma_{5}\tau^{a}\otimes i\gamma_{5}\tau^{a}\right)  ;\qquad
G_{V}\gamma_{\mu}\otimes\gamma_{\mu};\quad-G_{A}i\gamma_{\mu}\gamma_{5}%
\tau^{a}\otimes i\gamma_{\mu}\gamma_{5}\tau^{a}. \label{NJLnl}%
\end{equation}
In Eq.~(\ref{Lint}) $\overline{q}=(\overline{u},\overline{d})$ denotes the
quark flavor doublet field, $G_{i}$ are the four-quark coupling constants, and
$\tau^{a}$ are the Pauli isospin matrices. The separable nonlocal kernel of
the interaction determined in terms of form factors $f_{i}(x),$ with
normalization $f_{i}(0)=1,$ is motivated by instanton model of QCD vacuum. The
instanton model predicts the hierarchy of interactions in different channels.
It is most stronger in the pseudo-scalar and scalar channels providing the
spontaneous breaking of chiral symmetry. At the same time it is highly
suppressed in the vector and axial-vector channels. In these channels the
confinement force has to be taken into account in addition. Thus, in general
we treat differently the shape of form factors $f_{i}(x)$ in different channels.

In order to make gauge-invariant form of the nonlocal action with respect to
external gauge fields $V_{\mu}^{a}(x)$, we define in (\ref{Lint}) the
delocalized quark field, $Q(x)$ by using the Schwinger gauge phase factor
\begin{equation}
Q(x,y)=P\exp\left\{  i\int_{x}^{y}dz_{\mu}V_{\mu}^{a}(z)T^{a}\right\}
q(y),\qquad\overline{Q}(x,y)=Q^{\dagger}(x,y)\gamma^{0}, \label{Qxy}%
\end{equation}
where $P$ is the operator of ordering along the integration path, with $y$
denoting the position of the quark and $x$ being an arbitrary reference point.
The conserved vector and axial-vector currents in the scalar sector of the
model have been derived earlier in \cite{ADoLT00,DoBr03}. The extension of
these results onto the vector sector of the model is given in Appendix A.

The dressed quark propagator, $S(p)$, is defined as
\begin{equation}
S^{-1}(p)=\widehat{p}-M(p), \label{QuarkProp}%
\end{equation}
with the momentum-dependent quark mass found as the solution of the gap
equation%
\begin{equation}
M(p)=M_{q}f_{P}^{2}(p), \label{Mp}%
\end{equation}
where $f_{P}(p)$ is the normalized $4d$ Fourier transform of the $f_{P}(x)$.
The important property of the dynamical mass is that at low virtualities
passing through quark its mass is close to constituent mass, while at large
virtualities it goes to current mass value.

The quark-antiquark scattering matrix in different channels is found from the
Bethe-Salpeter equation as
\begin{equation}
\widehat{T}_{i}(q)=\frac{G_{i}}{1-G_{i}J_{i}(q)}, \label{ScattMatr}%
\end{equation}
with the polarization operator
\begin{equation}
J_{i}(q^{2})\delta_{ab}=-i\int\frac{d^{4}k}{\left(  2\pi\right)  ^{4}}%
f_{i}^{2}\left(  k\right)  f_{i}^{2}\left(  k+q\right)  Tr\left[
S(k)\Gamma_{i}^{a}S\left(  k+q\right)  \Gamma_{i}^{b}\right]  . \label{J}%
\end{equation}
The positions of mesonic bound states are determined as the poles of the
scattering matrix%
\begin{equation}
\left.  \det(1-G_{i}J_{i}(q))\right\vert _{q^{2}=m_{M}^{2}}=0. \label{PoleEq}%
\end{equation}
The quark-meson vertices in the pseudoscalar, vector and axial-vector channels
found from the residues of the scattering matrix are $\left(  k^{\prime
}=k+q\right)  $
\begin{align}
\Gamma_{\rho,s}^{a}\left(  k,k^{\prime}\right)   &  =g_{\rho qq}\gamma_{\mu
}\epsilon_{s}^{\mu}f^{V}(k)f^{V}(k^{\prime})\tau^{a},\qquad\Gamma_{\omega
,s}\left(  k,k^{\prime}\right)  =g_{\omega qq}\gamma_{\mu}\epsilon_{s}^{\mu
}f^{V}(k)f^{V}(k^{\prime}),\label{VectVertex}\\
\Gamma_{a_{1},s}^{a}\left(  k,k^{\prime}\right)   &  =g_{a_{1}qq}\gamma_{\mu
}\gamma_{5}\epsilon_{s}^{\mu}f^{V}(k)f^{V}(k^{\prime})\tau^{a},\quad
\Gamma_{\pi}^{a}\left(  k,k^{\prime}\right)  =\left(  g_{\pi qq}-\widetilde
{g}_{\pi qq}\widehat{q}/m_{\pi}\right)  i\gamma_{5}f^{P}(k)f^{P}(k^{\prime
})\tau^{a}\nonumber
\end{align}
with the quark-meson couplings found from
\begin{equation}
g_{Mqq}^{-2}=-\left.  \frac{dJ_{i}\left(  q^{2}\right)  }{dq^{2}}\right\vert
_{q^{2}=m_{M}^{2}}, \label{gM}%
\end{equation}
where $m_{\mathrm{M}}$ are physical masses of the $\rho\left(  \omega\right)
$- and $a_{1}$-mesons. Note, that the quark-pion vertex in (\ref{VectVertex})
takes into account effect of the $\pi-a_{1}$ mixing.

\section{Adler function within the nonlocal chiral quark model.}

\begin{figure}[h]
\hspace{2cm}\includegraphics[width=14.5cm]{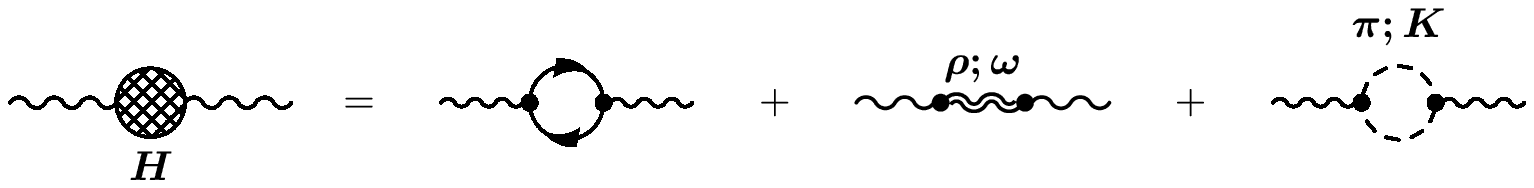}\caption{Schematic
representation of the vector polarization function (37). }%
\label{hpf}%
\end{figure}

Our goal is to obtain the vector current-current correlator and corresponding
Adler function by using the effective instanton-like model (\ref{Lint}) and
then to estimate the leading order hadron vacuum polarization correction to
muon anomalous magnetic moment $a_{\mu}$. In N$\chi$QM in the chiral limit the
(axial-)vector correlators have transverse character
\begin{equation}
\Pi_{\mu\nu}^{J}\left(  Q^{2}\right)  =\left(  g_{\mu\nu}-\frac{q^{\mu}q^{\nu
}}{q^{2}}\right)  \Pi_{J}^{\mathrm{N\chi QM}}\left(  Q^{2}\right)  ,
\label{PVmn}%
\end{equation}
where the polarization functions are given by the sum of the dynamical quark
loop, the intermediate (axial-)vector mesons and the higher order mesonic
loops contributions (see Fig. \ref{hpf})
\begin{equation}
\Pi_{J}^{\mathrm{N\chi QM}}\left(  Q^{2}\right)  =\Pi_{J}^{Q\mathrm{Loop}%
}\left(  Q^{2}\right)  +\Pi_{J}^{\mathrm{mesons}}\left(  Q^{2}\right)
+\Pi_{J}^{\chi\mathrm{Loop}}\left(  Q^{2}\right)  . \label{Pncqm}%
\end{equation}

The spectral representation of the polarization function consists
of zero width (axial-)vector resonances $\left(  \Pi_{J}^{\mathrm{mesons}%
}\left(  Q^{2}\right)  \right)  $ and two-meson states $\left(  \Pi_{J}%
^{\chi\mathrm{Loop}}\left(  Q^{2}\right)  \right)  .$ The dynamical quark loop
under condition of analytical confinement has no singularities in physical
space of momenta.

The dominant contribution to the vector current correlator at space-like
momentum transfer is given by the dynamical quark loop which was found in
\cite{DoBr03} with the result\footnote{Furthermore, the integrals over the
momentum are calculated by transforming the integration variables into the
Euclidean space, ($k^{0}\rightarrow ik_{4},$ $k^{2}\rightarrow-k^{2}$).}
\begin{align}
\Pi_{V}^{Q\mathrm{Loop}}\left(  Q^{2}\right)   &  =\frac{4N_{c}}{Q^{2}}%
\int\frac{d^{4}k}{\left(  2\pi\right)  ^{4}}\frac{1}{D_{+}D_{-}}\left\{
M_{+}M_{-}+\left[  k_{+}k_{-}-\frac{2}{3}k_{\perp}^{2}\right]  _{ren}\right.
\label{Ploop}\\
&  +\left.  \frac{4}{3}k_{\perp}^{2}\left[  \left(  M^{\left(  1\right)
}\left(  k_{+},k_{-}\right)  \right)  ^{2}\left(  k_{+}k_{-}-M_{+}%
M_{-}\right)  -\left(  M^{2}\left(  k_{+},k_{-}\right)  \right)  ^{\left(
1\right)  }\right]  \right\}  +\nonumber\\
&  +\frac{8N_{c}}{Q^{2}}\int\frac{d^{4}k}{\left(  2\pi\right)  ^{4}}%
\frac{M\left(  k\right)  }{D\left(  k\right)  }\left[  M^{\prime}\left(
k\right)  -\frac{4}{3}k_{\perp}^{2}M^{\left(  2\right)  }\left(
k,k+Q,k\right)  \right]  ,\nonumber
\end{align}
where the notations%

\[
k_{\pm}=k\pm Q/2,\qquad k_{\perp}^{2}=k_{+}k_{-}-\frac{\left(  k_{+}q\right)
\left(  k_{-}q\right)  }{q^{2}},\qquad D\left(  k\right)  =k^{2}+M^{2}(k),
\]%
\begin{equation}
M_{\pm}=M(k_{\pm}),\ \ \ \ \ D_{\pm}=D(k_{\pm}),
\end{equation}
are used. We also introduce the finite-difference derivatives defined for an
arbitrary function $F\left(  k\right)  $ as
\begin{equation}
F^{(1)}(k,k^{\prime})=\frac{F(k^{\prime})-F(k)}{k^{\prime2}-k^{2}},\qquad
F^{(2)}\left(  k,k^{\prime},k^{\prime\prime}\right)  =\frac{F^{(1)}%
(k,k^{\prime\prime})-F^{(1)}(k,k^{\prime})}{k^{\prime\prime2}-k^{\prime2}}.
\label{FDD}%
\end{equation}

\begin{figure}[h]
\includegraphics[width=14.5cm]{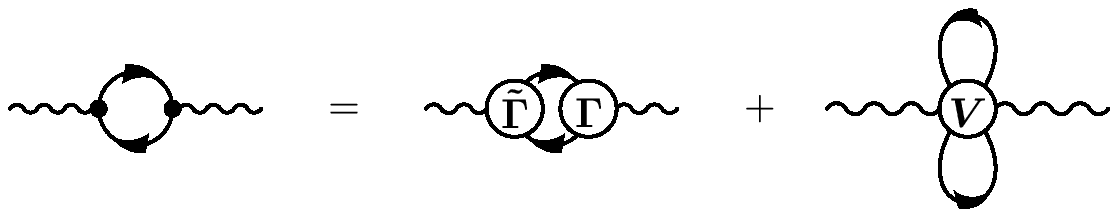}\caption{The dynamical quark-loop
contribution is the sum of dispersive and contact terms. In the dispersive
diagram $\widetilde{\Gamma}$ is the bare vertex and $\Gamma$ is the total
one.}%
\label{DispCont}%
\end{figure}

In (\ref{Ploop}) the first two lines represent the contribution of the
dispersive diagrams and the third line corresponds to the contact diagrams
(see Fig. \ref{DispCont} and ref. \cite{DoBr03} for details). The expression
for $\Pi_{V}^{Q\mathrm{Loop}}\left(  Q^{2}\right)  $ is formally divergent and
needs proper regularization and renormalization procedures which are
symbolically noted by $\left[  ..\right]  _{ren}$ for the divergent term. At
the same time the corresponding Adler function is well defined and finite.

Also we have checked that there is no pole in the vector correlator as
$Q^{2}\rightarrow0$, which simply means that photon remains massless with
inclusion of strong interaction. In the limiting cases the Adler function
derived from Eq. (\ref{Ploop}) satisfies general requirements of QCD (see
leading terms in (\ref{Dope}), (\ref{DpQCD}), and (\ref{D(0)}))%
\begin{equation}
A_{V}^{\mathrm{N\chi QM}}\left(  Q^{2}\rightarrow0\right)  =\mathcal{O}\left(
Q^{2}\right)  ,\qquad A_{V}^{\mathrm{N\chi QM}}\left(  Q^{2}\rightarrow
\infty\right)  =\frac{N_{c}}{12\pi^{2}}+\frac{O_{2}^{V}}{Q^{2}}+\mathcal{O}%
\left(  Q^{-4}\right)  . \label{Aasympt}%
\end{equation}
The leading high $Q^{2}$ asymptotics comes from the $\left[  k_{+}k_{-}%
-\frac{2}{3}k_{\perp}^{2}\right]  _{\mathrm{ren}}$ term in (\ref{Ploop}),
while the subleading asymptotics is driven by "tachionic" term with
coefficient \cite{DoBr03}%
\begin{equation}
O_{2}^{V}=-\frac{N_{c}}{2\pi^{2}}\int_{0}^{\infty}du\frac{uM\left(  u\right)
M^{\prime}\left(  u\right)  }{D\left(  u\right)  }. \label{Tachion}%
\end{equation}
In (\ref{Tachion}) and below we use the notations $u=k^{2}$ and $M^{\prime
}(u)=dM(u)/du$.

In the extended by vector interaction model (\ref{Lint}) one gets the
corrections due to the inclusion of $\rho$ and $\omega$ mesons which appear as
a result of quark-antiquark rescattering in these channels%
\begin{equation}
\Pi_{V}^{\mathrm{mesons}}\left(  Q^{2}\right)  =\frac{1}{2Q^{2}}\frac
{G_{V}B_{V}^{2}\left(  Q^{2}\right)  }{1-G_{V}J_{V}^{T}\left(  Q^{2}\right)
}, \label{PVmeson}%
\end{equation}
where $B_{V}\left(  q^{2}\right)  $ is the vector meson contribution to
quark-photon transition form factor%
\begin{equation}
B_{V}\left(  Q^{2}\right)  =8N_{c}i\int\frac{d^{4}k}{\left(  2\pi\right)
^{4}}\frac{f_{+}^{V}f_{-}^{V}}{D_{+}D_{-}}\left[  M_{+}M_{-}-k_{+}k_{-}%
+\frac{2}{3}k_{\perp}^{2}\left(  1-M^{2(1)}\left(  k_{+},k_{-}\right)
\right)  -\frac{4}{3}k_{\perp}^{2}\frac{f_{-}f^{(1)}(k_{-},k_{+})}{D_{-}%
}\right]  , \label{BV}%
\end{equation}
and $J_{V}^{T}\left(  q^{2}\right)  $ is the vector meson polarization
function defined in (\ref{J}) with $\Gamma_{\mu}^{T}=\left(  g_{\mu\nu}%
-q_{\mu}q_{\nu}/q^{2}\right)  \gamma_{\nu}$. As a consequence of the
Ward-Takahashi identity one has $B_{V}\left(  0\right)  =0$ as it should be.

To estimate the $\pi^{+}\pi^{-}$ and $K^{+}K^{-}$ vacuum polarization
insertions (chiral loops corrections) one may use the effective meson vertices
generated by the Lagrangian%
\begin{equation}
-ie\ A_{\mu}\left(  \pi^{+}\overleftrightarrow{\partial}_{\mu}\pi^{-}%
+K^{+}\overleftrightarrow{\partial}_{\mu}K^{-}\right)  \ . \label{EffInter}%
\end{equation}
By using the spectral density calculated from this interaction:
\begin{equation}
\rho_{V}^{\chi loop}\left(  t\right)  ={\frac{1}{12}}\left(  1-{\frac{4m_{\pi
}^{2}}{t}}\right)  ^{3/2}\Theta(t-4m_{\pi}^{2})+\left(  \pi\rightarrow
K\ \right)  , \label{SpectrDensChi}%
\end{equation}
one finds the contribution to the Adler function as%
\begin{equation}
D_{V}^{\chi\mathrm{Loop}}\left(  Q^{2}\right)  =\frac{1}{48\pi^{2}}\left[
a\left(  \frac{Q^{2}}{4m_{\pi}^{2}}\right)  +a\left(  \frac{Q^{2}}{4m_{K}^{2}%
}\right)  \right]  , \label{AChiLoop}%
\end{equation}
where
\begin{equation}
a\left(  t\right)  =\frac{1}{t}\left\{  3+t-\frac{3}{2}\sqrt{\frac{t+1}{t}%
}\left[  \operatorname{arctanh}\left(  \frac{1+2t}{2\sqrt{t\left(  t+1\right)
}}\right)  +i\frac{\pi}{2}\right]  \right\}  . \label{a(t)}%
\end{equation}
The estimate (\ref{AChiLoop}) of the chiral loop corrections corresponds to
the point-like mesons which becomes unreliable at large $t,$ where the meson
form factors has to be taken into account. This contribution corresponds to
the lowest order, $O(p^{4})$, calculations in $\chi$PT, is non-leading in the
formal $1/N_{c}$-expansion and provides numerically small addition. We avoid
to use literally the known two-loop, $O(p^{6})$, $\chi$PT result for the
chiral spectral function \cite{GolowKamb95,HoldLewM94}, since, as it was shown
there, the validity of the next-to-leading order in $p^{2}$ calculations is
justified only in the short interval of invariant masses $4m_{\pi}^{2}\leq
t\lesssim0.15$ GeV$^{2}.$ The higher-loop effects become important at higher momenta.

The resulting Adler function in N$\chi$QM is given by the sum of above
contributions
\begin{equation}
D_{V}\left(  Q^{2}\right)  =D_{V}^{Q\mathrm{Loop}}\left(  Q^{2}\right)
+D_{V}^{\mathrm{mesons}}\left(  Q^{2}\right)  +D_{V}^{\chi\mathrm{Loop}%
}\left(  Q^{2}\right)  . \label{Dncqm}%
\end{equation}

\section{Parameters of the extended model}

First of all we need to determine the shape of nonlocal form factors in the
kernel of the four-fermion interaction in (\ref{EffInter}). Within the
instanton model in the zero mode approximation the function $f_{p}(p^{2})$ is
expressed in terms of the modified Bessel functions. However, the screening
effect modifies the instanton shape at large distances leading to the
constraint instantons \cite{DEMM99}. To take into account screening and to
have also simpler analytical form for $f_{p}(p^{2})$ we shall use further the
Gaussian form for the instanton profile function
\begin{equation}
f_{P}(p)=\exp\left(  -p^{2}/\Lambda_{P}^{2}\right)  . \label{MassDyna}%
\end{equation}
Moreover, it is possible to show that for practical calculations of the
quantities that are defined in the space-like region the exact form of
nonlocality is not very important.

At the same time the profile in the (axial-)vector channels has not to be the
same as in the scalar channels. Indeed, in the instanton model in the zero
mode approximation there is strong interaction in the (pseudo-)scalar
channels, but there is no interaction at all in the (axial-)vector channels.
It means that the mechanism that bounds quarks in these states is different
from the one binding the light pseudoscalar mesons. The model also predicts
that while in the scalar channels the correlation length for nonlocality is
given by an instanton size, $\rho\approx0.3$ fm, the correlation length in the
vector channels is related to the distance between instanton and
antiinstanton, $R\approx1$ fm. Thus, it follows the expectation for the widths
of nonlocalities in momentum space in scalar and vector channels as
$\Lambda_{P}>>\Lambda_{V}$. In the present work we take the function with
property of analytical confinement as the nonlocal form factor in the vector
channels \cite{ADWB01,DoMKR}
\begin{equation}
f_{V}^{2}(u)=\frac{L_{V}\left(  u\right)  D\left(  u\right)  }{M_{q}%
^{2}\left(  u+L_{V}^{2}\left(  u\right)  \right)  },\qquad f_{V}(0)=1,
\label{fV}%
\end{equation}
where
\[
L_{V}\left(  u\right)  =\left(  \frac{u/\Lambda_{V}^{2}}{\exp(u/\Lambda
_{V}^{2})-1}\right)  ^{1/4}%
\]
and $\Lambda_{V}$ is the momentum space width of nonlocality in the vector
channel. The function (\ref{fV}) has the property that it tends to zero at
large positive values of $u$ and has no poles.

The N$\chi$QM can be viewed as an approximation of large-$N_{c}$ QCD where the
only new interaction terms, retained after integration of the high frequency
modes of the quark and gluon fields down to a nonlocality scale $\Lambda$ at
which spontaneous chiral symmetry breaking occurs, are those which can be cast
in the form of four-fermion ope\-ra\-tors (\ref{Lint}), (\ref{NJLnl}). The
parameters of the model are then the nonlocality scales $\Lambda$ and the
four-fermion coupling constants $G$.
%These couplings can be traded for the mass M_{Q} of the constituent chiral quark,
%which appears as a non-trivial solution to the gap equation involving G_{P}.

The parameters of the model are fixed in a way typical for effective
low-energy quark models. In quark models one usually fits the pion decay
constant, $f_{\pi}$, to its experimental value, which in the chiral limit
reduces to $86$ \textrm{MeV} \cite{LeutG}. In N$\chi$QM extended by vector
interactions the constant, $f_{\pi}$, is determined by
\begin{equation}
f_{\pi}^{2}=\frac{N_{c}}{4\pi^{2}}\int\limits_{0}^{\infty}du\ u\frac
{M^{2}(u)-uM(u)M^{\prime}(u)+u^{2}M^{\prime}(u)^{2}}{D^{2}\left(  u\right)
}+\frac{G_{A}j_{AP}^{2}\left(  0\right)  }{1-G_{A}J_{A}^{L}\left(  0\right)
}, \label{Fpi2_M}%
\end{equation}
where%
\begin{equation}
J_{A}^{L}\left(  0\right)  =\frac{N_{c}}{2\pi^{2}}\int\limits_{0}^{\infty
}du\ uf_{V}^{4}\left(  u\right)  \frac{M^{2}(u)-u/2}{D^{2}\left(  u\right)
},\qquad j_{AP}\left(  0\right)  =-\frac{N_{c}}{2\pi^{2}}\int\limits_{0}%
^{\infty}du\ uf_{V}^{4}\left(  u\right)  \frac{M(u)-uM^{\prime}(u)/2}%
{D^{2}\left(  u\right)  }. \label{JL}%
\end{equation}
The second term in (\ref{Fpi2_M}) arises due to the $\pi-a_{1}$ mixing effect.
In the local NJL model the $\pi-a_{1}$ mixing plays important role and leads
to large corrections to observables of order $\sim30\%$. However, in the
nonlocal models the mixing becomes a small effect and this is a general
property of such models. For example, it corrects the value of $f_{\pi}$ at
the level of $\sim1\%$.

The couplings $G_{V}^{\rho,\omega}$ and $G_{A}^{a_{1}}$ are fixed by requiring
that poles of the scattering matrix (\ref{PoleEq}) coincide with physical
meson masses $\left(  m_{\rho}=770\text{ MeV, }m_{\omega}=783\text{ MeV,
}m_{a_{_{1}}}=1230\text{ MeV}\right)  $. The parameter $\Lambda_{V}$ is chosen
to fit the widths of the $\rho\rightarrow\pi\pi$ and $\rho\rightarrow
e^{+}e^{-}$ decays (see details in Appendix B). One gets the values of the
model parameters%
\begin{align}
M_{q}  &  =0.24~\mathrm{GeV,}\qquad\Lambda_{P}=1.11~\mathrm{GeV,\qquad}%
\Lambda_{V}=0.3~\mathrm{GeV,}\label{Lambda's}\\
G_{P}^{\pi}  &  =27.4~\mathrm{GeV}^{-2},\qquad G_{V}^{\rho}=-1.96~\mathrm{GeV}%
^{-2},\qquad G_{V}^{\omega}=-1.78~\mathrm{GeV}^{-2},\qquad G_{A}^{a_{1}%
}=-0.03~\mathrm{GeV}^{-2}. \label{G's}%
\end{align}

%In particular, one obtains
%\begin{equation}
%\langle\bar{q}q\rangle=-\left(  214~\mathrm{MeV}\right)  ^{3},\qquad O_{2}%
%^{V}=0.003\quad\mathrm{GeV}^{2}, \label{qq}%
%\end{equation}
%where the matrix element $O_{2}^{V}$ of the dimension 2 operator is defined in
%(\ref{Tachion}) and the quark condensate is given by
%\begin{equation}
%\langle\bar{q}q\rangle=-\frac{N_{c}}{4\pi^{2}}\int du\ u\frac{M(u)}{D\left(
%u\right)  }. \label{QQcond}%
%\end{equation}

It is important to note that within the N$\chi$QM one gets the ratio $G_{P}%
\gg\left\vert G_{V,A}\right\vert $. This is opposite to the local NJL model
where one has $G_{P}\approx5~\mathrm{GeV}^{-2}$ and $\left\vert G_{V}%
\right\vert \approx10~\mathrm{GeV}^{-2}$. In \cite{YaZakh} it was noted that
the large value of $\left\vert G_{V}\right\vert $ leads to strong
contradiction with QCD sum rules results in the $\rho$ channel. Moreover, in
\cite{PPdR98} it was noted that there is no overlap between applicability
regions of NJL and OPE QCD. It is clear from (\ref{G's}) that within the
N$\chi$QM the vector meson corrections become much smaller thus resolving the
problem. Note also that the ratio of the widths $\Lambda_{P}$ and $\Lambda
_{V}$ is in accordance with the instanton liquid model prediction.

\begin{figure}[h]
\includegraphics[height=7.5cm]{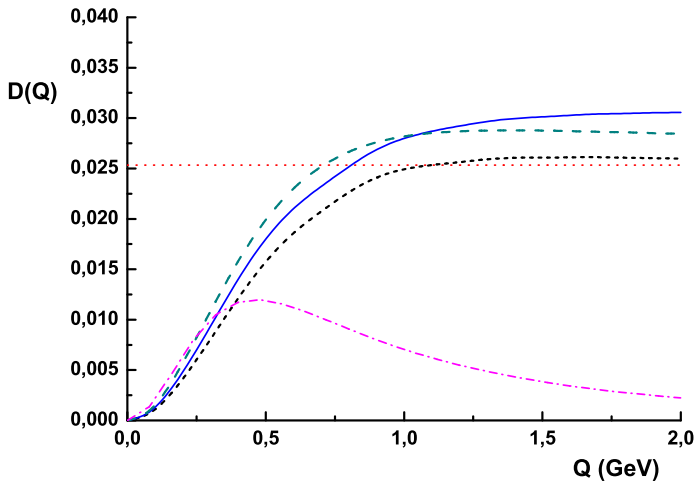}\caption{The Adler function from
the N$\chi$QM contributions: dynamical quark loop (short dashed), quark +
chiral loops + vector mesons (full line) versus the ALEPH data (dashed). The
dash-dotted line is the prediction of the constituent quark model (ENJL) and
the dotted line is the asymptotic freedom prediction, $1/4\pi^{2}$.}%
\label{AdlerV}%
\end{figure}

With the above set of parameters the Adler function in the vector channel
calculated in N$\chi$QM (\ref{Dncqm}) is presented in Fig. \ref{AdlerV} and
the model estimate for the hadronic vacuum polarization to $a_{\mu}$ given by
(\ref{aAd}) is
\begin{equation}
a_{\mu}^{\mathrm{hvp}~\left(  1\right)  ;\mathrm{N\chi QM}}=6.2\left(
0.4\right)  \cdot10^{-8}\ ,\label{AMMncqm}%
\end{equation}
where the various contributions to $a_{\mu}^{\mathrm{hvp}~\left(  1\right)
;\mathrm{N\chi QM}}$ are%
\begin{equation}
a_{\mu}^{\mathrm{hvp}~\left(  1\right)  ;\mathrm{Qloop}}=5.3\cdot10^{-8},\quad
a_{\mu}^{\mathrm{hvp}~\left(  1\right)  ;\mathrm{Vmesons}}=0.1\cdot
10^{-8},\quad a_{\mu}^{\mathrm{hvp}~\left(  1\right)  ;\mathrm{\chi Loop}%
}=0.8\cdot10^{-8}%
\end{equation}
and the error in (\ref{AMMncqm}) is due to incomplete knowledge of the higher
order effects in nonchiral corrections. One may conclude, that the agreement
of the N$\chi$QM estimate with the phenomenological determinations is rather
good. With the same model parameters one gets the estimate for the $\alpha
^{2}$ hadronic contribution to the $\tau$-lepton anomalous magnetic moments
\begin{equation}
a_{\tau}^{\mathrm{hvp}~\left(  1\right)  ;\mathrm{N\chi QM}}=
3.1\left(0.2\right)  \cdot10^{-6}\ ,\label{AMMtauNCQM}%
\end{equation}
which is in agreement with phenomenological determination
\begin{equation}
a_{\tau}^{\mathrm{hvp}~\left(  1\right)  ;\mathrm{exp}}=\left\{
\begin{array}
[c]{c}%
3.383\left(  0.111\right)  \cdot10^{-6},\qquad\cite{Jegerlehner96},\\
3.536\left(  0.038\right)  \cdot10^{-6},\qquad\cite{Narison01}.
\end{array}
\right.  \ \label{AMMtauEXP}%
\end{equation}

\section{Other model approaches to Adler function and $a_{\mu}^{\mathrm{hvp}%
~\left(  1\right)  }$.}

In \cite{BijdeRaf94} the vector two-point function has been calculated in the
extended Nambu--Jona-Lasinio (ENJL) model with the result
\begin{equation}
\Pi_{V}^{ENJL}(Q^{2})={\frac{\overline{\Pi}_{V}(Q^{2})}{1+Q^{2}{\frac{8\pi
^{2}G_{V}}{3\Lambda_{r}^{2}}}\overline{\Pi}_{V}(Q^{2})}}\ , \label{Pnjl}%
\end{equation}
where $\overline{\Pi}_{V}(Q^{2})$ is given by a loop of constituent massive
quarks as
\begin{equation}
\overline{\Pi}_{V}(Q^{2})=\frac{3}{2\pi^{2}}\int_{0}^{1}dxx\left(  1-x\right)
\Gamma\left(  0,x_{Q}\right)  , \label{Bnjl}%
\end{equation}
with $x_{Q}=\left(  M_{Q}^{2}+Q^{2}x\left(  1-x\right)  \right)  /\Lambda
_{r}^{2}$ and $\Gamma\left(  n,x\right)  $ is the incomplete Gamma function.
The parameters of the model are the constituent quark mass $M_{Q}=265$ MeV and
the ultraviolet regulator of the model $\Lambda_{r}=1.165$ GeV. The Adler
function predicted by ENJL model is presented in Fig. \ref{AdlerV}. It is
clear from the figure that the ENJL model with constituent quarks does not
interpolate correctly the transition from low to high momenta and fails to
describe the hadronic data starting already at low momenta.Within ENJL the
estimate of $a_{\mu}^{\mathrm{hvp}~\left(  1\right)  }$ was found as
\cite{deRaf93}
\begin{equation}
\left[  a_{\mu}^{\mathrm{hvp}~\left(  1\right)  }\right]  _{\mathrm{ENJL}%
}\simeq7.5\cdot10^{-8}\ . \label{AMM1njl}%
\end{equation}

Attempt to improve the situation has been done in \cite{PPdR98} by introducing
into the model infinite set of higher dimension terms. Effectively this
procedure reduce to delocalization of the quark-quark interaction, the
property inherent to the instanton-like models. Improved version of the ENJL
is close to the predictions of the minimal hadronic approximation model
(MHA)~\cite{PPdR98} based on the local duality. Assuming that the spectral
density $\rho_{V}(t)$ is given by a sum ansatz of a single, zero width vector
meson resonance and the QCD perturbative continuum contribution one has%
\begin{equation}
\frac{1}{4\pi^{2}}\rho_{V}^{MHA}\left(  t\right)  =2f_{V}^{2}M_{V}^{2}%
\delta\left(  t-M_{V}^{2}\right)  +\frac{N_{c}}{12\pi^{2}}\theta\left(
t-s_{0}\right)  ,\qquad\label{PVmha}%
\end{equation}
and the Adler function in the Euclidean region is given by
\begin{equation}
A_{V}^{MHA}\left(  Q^{2}\right)  =2f_{V}^{2}M_{V}^{2}\frac{Q^{2}}{\left(
Q^{2}+M_{V}^{2}\right)  ^{2}}+\frac{N_{c}}{12\pi^{2}}\frac{Q^{2}}{Q^{2}+s_{0}%
}, \label{Amha}%
\end{equation}
with $s_{0}\approx1\div1.5$ GeV$^{2}$ being the continuum threshold. This
model predicts correct asymptotic behaviour of the Adler function. By taking
the model parameters $M_{V}=0.750~GeV,$ $f_{V}^{2}=2f_{\pi}^{2}M_{V}^{2},$
$s_{0}\approx1.35\quad\mathrm{GeV}$ one finds an estimate for $a_{\mu
}^{\mathrm{hvp}~\left(  1\right)  }$ as \cite{Knecht}%
\begin{equation}
\left[  a_{\mu}^{\mathrm{hvp}~\left(  1\right)  }\right]  _{\mathrm{MHA}%
}=\left(  5.7\pm1.7\right)  \cdot10^{-8}\ . \label{AM1mha}%
\end{equation}

We also wish to recall that some time ago the hadronic contribution to the
photon vacuum polarization to $a_{\mu}^{\mathrm{hvp}~\left(  1\right)  }$ has
been estimated in the gauged nonlocal constituent quark model (GNC), the
approach which is quite similar to the present model, with the result
\cite{HoldLewM94}%
\begin{equation}
\left[  a_{\mu}^{\mathrm{hvp}~\left(  1\right)  }\right]  _{\mathrm{GNC}%
}=\left(  6.3\pm0.5\right)  \cdot10^{-8}\ . \label{AM1holdom}%
\end{equation}
It was also found that the model with momentum dependent quark mass give more
favorable results with respect to the constituent quark model with constant
masses. The GNC result is a sum of the dynamical quark loop contribution and
of the one- and two-meson loops contributions. We think, however, that the
two-loop contribution found in \cite{HoldLewM94}, which is almost the same as
the one-loop result, is overestimated in the model calculations. As we already
noted above the two-loop results have very narrow region of applicability.

Good agreement with the vector Adler function extracted from the experiment
has been reached in \cite{SolMilt} by using the analytic perturbative (AP)
approach \cite{AnalPT}. Within this model the perturbative expansion of the
Adler function valid at high energy analytically continued to the infrared
region where regular behaviour is predicted. There are two parameters in this
model: the QCD parameter $\Lambda_{QCD}=420$ MeV and the light quark masses
$m_{u,d}=250$ MeV. Note, that like to N$\chi$QM considerations the AP approach
prefers to use rather low values of the quark masses to describe
quantitatively the Adler function.

Finally we note that first evaluations of $a_{\mu}^{\mathrm{hvp}~}$ on the
lattice appeared recently \cite{Latt}%
\begin{equation}
\left[  a_{\mu}^{\mathrm{hvp}~}\right]  _{\mathrm{Lattice}}=\left(
4.46\pm0.23\right)  \cdot10^{-8}. \label{AMMlattice}%
\end{equation}
Still this calculations contains rather big uncertainties and the above error
does not take into account large systematical errors from the quenching
approximation, unphysically large quark masses and finite volume effects.

\section{$V-A$ correlator}

For consistency we also give the expressions for the difference of the $V$ and
$A$ correlators (\ref{Pncqm}) in the extended N$\chi$QM and discuss the
related chiral sum rules. The axial currents correlator is given by the sum of
the dynamical quark loop, the intermediate axial-vector mesons propagators and
the meson chiral loops (see some details in Appendix C). The dynamical quark
loop contribution to the $V-A$ correlator has been found earlier in
\cite{DoBr03} and reads
\begin{align}
\Pi_{V-A}^{\mathrm{Loop}}\left(  Q^{2}\right)   &  =-\frac{4N_{c}}{Q^{2}}%
\int\frac{d^{4}k}{\left(  2\pi\right)  ^{4}}\frac{1}{D_{+}D_{-}}\left\{
M_{+}M_{-}+\frac{4}{3}k_{\perp}^{2}\left[  -\sqrt{M_{+}M_{-}}M^{\left(
1\right)  }\left(  k_{+},k_{-}\right)  +\right.  \right. \label{VmAmodel}\\
&  \left.  \left.  +\left(  \sqrt{M}^{\left(  1\right)  }\left(  k_{+}%
,k_{-}\right)  \right)  ^{2}\left(  \sqrt{M_{+}}k_{+}+\sqrt{M_{-}}%
k_{-}\right)  ^{2}\right]  \right\}  .\nonumber
\end{align}
The integrand of the above expression is positive-definite in accordance with
the Witten inequality.

In the extended model one gets the corrections from the vector and
axial-vector intermediate mesons generated via the quark-quark rescattering to
the $V-A$ polarization function%
\begin{equation}
\Pi_{V-A}^{\mathrm{meson}}\left(  Q^{2}\right)  =\Pi_{V}^{\mathrm{meson}%
}\left(  Q^{2}\right)  -\Pi_{A}^{\mathrm{meson}}\left(  Q^{2}\right)  ,
\label{PVmAmesons}%
\end{equation}
where $\Pi_{V}^{\mathrm{meson}}\left(  Q^{2}\right)  $ is defined in
(\ref{PVmeson}) and the axial-vector polarization function due to the $a_{1}%
$-meson is%
\begin{equation}
\Pi_{A}^{\mathrm{meson}}\left(  Q^{2}\right)  =-\frac{1}{2}\frac{G_{A}%
B_{A}^{2}\left(  Q^{2}\right)  }{1-G_{A}J_{A}^{T}\left(  Q^{2}\right)  },
\label{PAmeson}%
\end{equation}
where $B_{A}\left(  q^{2}\right)  $ is the transition form factor of the
axial-vector current to quark-antiquark pair
\begin{align}
B_{A}\left(  Q^{2}\right)   &  =8N_{c}i\int\frac{d^{4}k}{\left(  2\pi\right)
^{4}}\frac{f_{+}^{V}f_{-}^{V}}{D_{+}D_{-}}\left\{  -M_{+}M_{-}-k_{+}%
k_{-}+\frac{2}{3}k_{\perp}^{2}\left[  1-M^{(1)}\left(  k_{+},k_{-}\right)
\left(  \sqrt{M_{+}}-\sqrt{M_{-}}\right)  ^{2}\right]  -\right. \label{BA}\\
&  \left.  -\frac{4}{3}k_{\perp}^{2}\frac{f_{-}f^{(1)}(k_{-},k_{+})}{D_{-}%
}\right\}  ,\nonumber
\end{align}
and $J_{A}^{T}\left(  Q^{2}\right)  $ is defined in (\ref{J}). At zero
momentum $B_{A}\left(  0\right)  \neq0$ in accordance with the effect of
$\pi-a_{1}$ meson mixing effect. As a consequence the total $V-A$ correlator
is consistent with the first Weinberg sum rule: $\lim_{Q^{2}\rightarrow0}%
Q^{2}\Pi_{V-A}^{T}\left(  Q^{2}\right)  =-f_{\pi}^{2}$, where $f_{\pi}^{2}$ is
defined in (\ref{Fpi2_M}). Due to very small value of $G_{A}$ the effect of
the $a_{1}$ meson on the axial-vector correlator is very tiny.

Let us now consider the low-energy region where the effective model
(\ref{Lint}) should be fully predictive. From (\ref{VmAmodel}) and the
Das-Guralnik-Mathur-Low-Yuong (DGMLY) sum rule \cite{DGMLY}, $\left(  \Delta
m_{\pi}^{2}\equiv m_{\pi^{\pm}}^{2}-m_{\pi^{0}}^{2}\right)  $%
\begin{equation}
-\frac{1}{4\pi^{2}}\int_{0}^{s_{0}\rightarrow\infty}dss\ln\frac{s}{\mu^{2}%
}\left[  \rho_{V}\left(  s\right)  -\rho_{A}\left(  s\right)  \right]
=\int_{0}^{\infty}dQ^{2}\left[  -Q^{2}\Pi_{V-A}^{T}\left(  Q^{2}\right)
\right]  =\frac{4\pi f_{\pi}^{2}}{3\alpha}\Delta m_{\pi}^{2}, \label{DGMLY}%
\end{equation}
we estimate the electromagnetic pion mass difference to be $\left(  \Delta
m_{\pi}\equiv m_{\pi^{\pm}}-m_{\pi^{0}}\right)  $
\begin{equation}
\left[  \Delta m_{\pi}\right]  _{N\chi QM}=4.2\mathrm{~MeV}, \label{dMmodel}%
\end{equation}
which is in remarkable agreement with the experimental value (after
subtracting the $m_{d}-m_{u}$ effect) \cite{GasLeut85}
\begin{equation}
\left[  \Delta m_{\pi}\right]  _{\mathrm{exp}}=4.43\pm0.03\mathrm{~MeV}.
\label{dMpiExp}%
\end{equation}
The $\pi^{+}-\pi^{0}$ electromagnetic mass difference is another observable
which offers the possibility to test the quality of the matching between
long-distance and short-distance behaviours.

With help of the Das-Mathur-Okubo (DMO) sum rule \cite{DMO},%
\begin{equation}
I_{\mathrm{DMO}}\left(  s_{0}\rightarrow\infty\right)  =\frac{1}{4\pi^{2}}%
\int_{0}^{s_{0}\rightarrow\infty}\frac{ds}{s}\left[  \rho_{V}\left(  s\right)
-\rho_{A}\left(  s\right)  \right]  =\left.  \frac{\partial}{\partial Q^{2}%
}\left[  Q^{2}\Pi_{V-A}^{T}\left(  Q^{2}\right)  \right]  \right\vert
_{Q^{2}\rightarrow0}=\frac{1}{3}f_{\pi}^{2}\left\langle r_{\pi}^{2}%
\right\rangle -F_{A}, \label{DMO}%
\end{equation}
where $F_{A}$ is the pion axial-vector form factor and $\left\langle r_{\pi
}^{2}\right\rangle $ is the electromagnetic pion radius squared, we estimate
the electric polarizability of the charged pions by using \cite{Geras79}
\begin{equation}
\alpha_{\pi^{\pm}}^{E}=\frac{\alpha}{m_{\pi}}\left[  \frac{\left\langle
r_{\pi}^{2}\right\rangle }{3}-\frac{I_{\mathrm{DMO}}}{f_{\pi}^{2}}\right]  .
\label{PiPolariz}%
\end{equation}

With the experimental value for the pion mean squared radius $\left\langle
r_{\pi}^{2}\right\rangle =\left(  0.439\pm0.008\right)  $ fm$^{2}$ \cite{PDG}
and the value of the $I_{DMO}$ integral estimated from the OPAL data
\cite{OPAL}
\begin{equation}
\left[  I_{\mathrm{DMO}}\left(  s_{0}=m_{\tau}^{2}\right)  \right]
_{\mathrm{exp}}=\left(  26.3\pm1.8\right)  \cdot10^{-3}%
\end{equation}
one gets from Eq. (\ref{PiPolariz}) the result \cite{OPAL}
\begin{equation}
\left[  \alpha_{\pi^{\pm}}^{E}\right]  _{\mathrm{exp}}^{\mathrm{OPAL}}=\left(
2.71\pm0.88\right)  \cdot10^{-4}\mathrm{~fm}^{3}. \label{PiPolarizOPAL}%
\end{equation}
Another experimental estimate of the pion polarizability follows from recent
measurments by PIBETA collaboration \cite{PIBETA} of the pion axial-vector
form factor $F_{A}$ with a result $F_{A}=0.443(15)\cdot F_{V}$ (full data set)
and $F_{A}=0.480(16)\cdot F_{V}$ (kinematically restricted data set). Then
using the Terentyev relation \cite{Teren72}
\begin{equation}
\alpha_{\pi^{\pm}}^{E}=\frac{\alpha F_{A}}{8\pi^{2}m_{\pi}f_{\pi}^{2}}.
\label{PiPolarizFA}%
\end{equation}
one yields \cite{PIBpoc}
\begin{equation}
\left[  \alpha_{\pi^{\pm}}^{E}\right]  _{\mathrm{exp}}^{\mathrm{PIBETA}%
}=\left\{
\begin{array}
[c]{l}%
2.68(9)\cdot10^{-4}\mathrm{~fm}^{3}\qquad\mathrm{{full\ data\ set},}\\
2.90(9)\cdot10^{-4}\mathrm{~fm}^{3}\qquad
\mathrm{kinematically\ restricted\ data\ set}.
\end{array}
\right.  \label{PiPolarizPIBETA}%
\end{equation}

Within N$\chi$QM one obtains the estimates of the low energy constants
\begin{equation}
\left[  I_{\mathrm{DMO}}\right]  _{\mathrm{N\chi QM}}=23\cdot10^{-3}%
,\qquad\left[  \left\langle r_{\pi}^{2}\right\rangle \right]  _{\mathrm{N\chi
QM}}=0.435\mathrm{~fm}^{2}, \label{L10m}%
\end{equation}
where these values are derived by calculating the derivatives of $\Pi
^{V-A}\left(  Q^{2}\right)  $ and the electromagnetic pion form factor at zero
momentum, with vector meson \cite{DoMKR} and chiral loop corrections being
included (see Appendix). By using the values given in Eqs. (\ref{L10m}) we
find from (\ref{PiPolariz}) the value
\begin{equation}
\left[  \alpha_{\pi^{\pm}}^{E}\right]  _{N\chi QM}=2.9\cdot10^{-4}%
\mathrm{fm}^{3}, \label{AlMod}%
\end{equation}
which is close to experimental numbers (\ref{PiPolarizOPAL}) and
(\ref{PiPolarizPIBETA}). Thus, we see that the model prediction for the pion
polarizability, Eq.~(\ref{AlMod}), is in a very reasonable agreement with the
experimental data.

\begin{figure}[h]
\includegraphics[height=7.5cm]{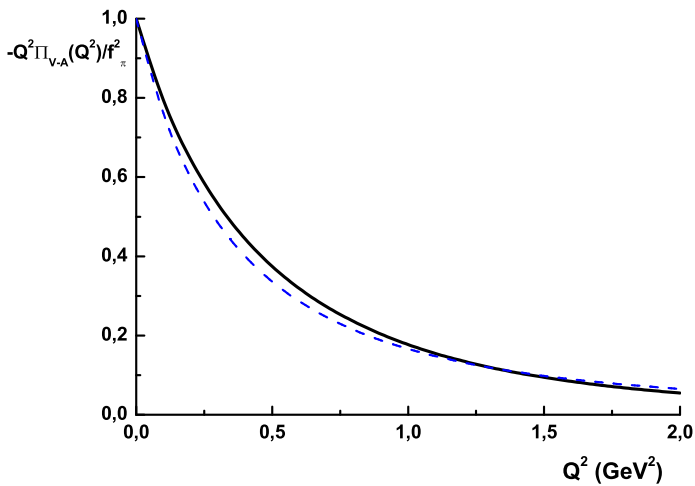}\caption{Normalized $V-A$
correlation function constructed in the N$\chi$QM (solid line) and
reconstructed from the ALEPH experimental spectral function (dashed line).}%
\label{Fv-a}%
\end{figure}

Next, we compare the $V-A$ correlator predicted in N$\chi$QM with the ALEPH
data. As above, we use $s_{0}=2.5$ GeV$^{2}$ as an upper integration limit,
the value at which all chiral sum rules are satisfied assuming that
$\rho_{V-A}(s)=0$ at $s\geq s_{0}$, (\ref{IIWSR}). Finally, a kinematic pole
at $q^{2}=0$ is added to the axial-vector spectral function. The resulting
unsubtracted dispersion relation between the measured spectral densities and
the correlation functions becomes
\begin{equation}
\Pi_{V-A}^{T}\left(  Q^{2}\right)  =\frac{1}{4\pi^{2}}\int_{0}^{s_{0}}%
ds\frac{\rho_{V}\left(  s\right)  -\rho_{A}\left(  s\right)  }{s+Q^{2}}%
-\frac{f_{\pi}^{2}}{Q^{2}}, \label{VmAreconstr}%
\end{equation}
where $f_{\pi}^{2}$ is given by the first Weinberg sum rule
\begin{equation}
f_{\pi}^{2}=\frac{1}{4\pi^{2}}\int_{0}^{s_{0}}ds\left[  v_{1}(s)-a_{1}%
(s)\right]  .
\end{equation}
Having transformed the data into the Euclidean space, we may now proceed with
the comparison to the model, which obviously applies to the Euclidean domain
only. The resulting normalized $V-A$ correlation functions corresponding to
the experimental data and the N$\chi$QM prediction are shown in Fig.
\ref{Fv-a}.

\section{Conclusions}

In this work we have analyzed the vector Adler function for Euclidean
(spacelike) momenta within an effective nonlocal chiral quark model motivated
by the instanton model of QCD vacuum. To this end, we have derived the
conserved vector and axial-vector currents and have constructed the
Euclidean-momentum correlation functions of the vector and axial-vector
currents in the extended by inclusion of vector mesons degrees of freedom
model. The dominant contributions to the polarization functions and to the
corresponding Adler functions come from the loop contributions of the light
dynamical quarks. It is this contribution that provides the matching between
low energy hadronized phase and the high energy QCD which is clearly seen in
the behaviour of the Adler function (Figs. \ref{AdlerALEPH} and \ref{AdlerV}).
The results obtained are close to estimates of the vector Adler function and
the $V-A$ correlator extracted directly from the ALEPH data on hadronic
inclusive $\tau$ decays and transformed by dispersion relations to the
spacelike region.

We further use the vector Adler function to calculate the $\alpha^{2}$
hadronic contribution to the muon anomalous magnetic moment, $a_{\mu
}^{\mathrm{hvp}~\left(  1\right)  }$, and found the value which is in
reasonable agreement with the latest precise phenomenological numbers. The
main reason why we discuss the N$\chi$QM is that it offers the possibility to
go beyond the leading $O(p^{6})$ contribution of $\chi$PT. This will appear to
be a crucial issue in consideration of the hadronic light-by-light
contribution. Reproducing the phenomenological determination of $a_{\mu
}^{\mathrm{hvp}~\left(  1\right)  }$, it becomes possible to make in future
reliable estimates of $a_{\mu}^{\mathrm{hvp}~\left(  2\right)  }$ and $a_{\mu
}^{\mathrm{h.~L\times L}}.$

In the N$\chi$QM extended by inclusion of vector and axial-vector mesons we
show that the influence of these states on the Adler function and some
low-energy observables is very small, at the level of few percents. This is
because the physical states corresponding to these channels are rather heavy
and, as consequence, have the couplings of quark-quark interactions much
smaller than in the (pseudo-)scalar channels.

We also estimated the contributions to the correlators and the low energy
observables of the nonchiral corrections, formally suppressed in $1/N_{c}$
expansion, by using the one loop results of the chiral perturbative theory.
These corrections are typically of order $10-20\%$ comparing to the dynamical
quark contributions. In this way the form factor dependence of the currents
with mesons is not taken into account and predictions become untrustable at
higher momenta. In perspective we plan to cure this disease and also take into
account effectively the perturbative gluon corrections which become important
at higher momenta.

From the properties of the $V-A$ correlator we have shown the fulfillment of
the low-energy relations. The values of the $\pi^{\pm}-\pi^{0}$
electromagnetic mass difference and the electric pion polarizability are
estimated and found to be close to the experimental values. We stress that the
momentum dependence of the dynamical quark mass is crucial for the reproducing
the empirical Adler function. The $V-A$ combination receives no contribution
from perturbative effects and provides a clean probe for chiral symmetry
breaking and a test ground for model verification.

Finally, we would like to note that the effective models based on the
underlying symmetries of strong interactions are usually operative at low
energies and fails in description of the high and even intermediate energy
regions. An essential advantage of the nonlocal models is that they provide
correct interpolation between the high energy behaviour where they are reduced
to asymptotically free model and the low energy behaviour where they share the
chiral symmetry and its spontaneous breaking. This is because the main
elements of diagram technics, the quark propagator and the quark couplings
with currents, become local at high virtualities. This property allows us to
straightforwardly reproduce the leading terms of the operator product
expansion. For instance, the correct high energy asymptotics are derived for
the $V-A$ correlator (second Weinberg sum rule) \cite{Bron99,DoBr03}, the
Adler function (this work), the topological susceptibility \cite{DoBr03}, the
pion transition form factor \cite{AD02} and some other quantities
\cite{DoLT98}, while in all these cases the constituent quark model, with
massive momentum independent quark masses, fails to explain the asymptotics.
The nonlocality originated from the quark-quark interactions due to exchange
of instantons also is of great importance in describing the hadronic
distribution functions in order to produce the correct end-point behavior as
it was shown for the case of the pion distribution amplitude \cite{D95} and
structure function \cite{DoLT98}.

The author is grateful to W. Broniowski, S. B. Gerasimov, N. I. Kochelev, P.
Kroll, S.V. Mikhailov, A. E. Radzhabov, O.P. Solovtsova and M. K. Volkov for
useful discussions on the subject of the present work. AED thanks for partial
support from RFBR (Grants nos. 02-02-16194, 03-02-17291, 04-02-16445), INTAS
(Grant no. 00-00-366). We are grateful to the Heisenberg - Landau program for support.

\appendix

\section{Conserved vector and axial-vector currents}

The Noether currents and the corresponding vertices are formally obtained as
functional derivatives of the action (\ref{Lint}) with respect to the external
fields at zero value of the fields. For our purpose, it is necessary to
construct the quark-current vertices that involve one or two currents (contact
terms). In the presence of the nonlocal interaction the conserved currents
include both local and nonlocal terms. The technique of expansion of the
path-ordered exponent in the external fields and derivation of the conserved
currents has been reviewed in \cite{ADoLT00,DoBr03}. Here we present the
extension of the previuos results for the model (\ref{Lint}) that includes in
addition the vector and axial-vector mesons degrees of freedom.

The (axial-)vector meson contribution to the bare (axial-)vector vertex
obtained by the differentiation the action (\ref{Lint}) with respect to the
external (axial-)vector field is given by the formula
\begin{equation}
\Delta\widetilde{\Gamma}_{\mu}^{(5)a}(p,q,p^{\prime}=p+q)=T^{a}\left\{
-G_{J}f^{V}(p^{\prime})f^{V}(p)\left[  \left(  g_{\mu\nu}-\frac{q^{\mu}q^{\nu
}}{q^{2}}\right)  \gamma_{\nu}\widetilde{B}\left(  q^{2}\right)  +\frac
{q^{\mu}\widehat{q}}{q^{2}}\widetilde{C}\left(  q^{2}\right)  \right]
\right\}  \left(  \gamma_{5}\right)  , \label{GVbare}%
\end{equation}
where $q$ is the momentum corresponding to the current, and $p$ $(p^{\prime})$
is the incoming (outgoing) momentum of the quark,%
\begin{align}
\widetilde{B}\left(  q^{2}\right)   &  =16N_{c}i\int\frac{d^{4}k}{\left(
2\pi\right)  ^{4}}\frac{2}{3}k_{\perp}^{2}\frac{f^{V}\left(  k\right)
f^{V(1)}(k,k+q)}{D\left(  k\right)  },\label{BVb}\\
\widetilde{C}\left(  q^{2}\right)   &  =16N_{c}i\int\frac{d^{4}k}{\left(
2\pi\right)  ^{4}}\left(  kq\right)  \frac{f^{V}\left(  k\right)  f^{V}%
(k+q)}{D\left(  k\right)  }. \label{CVb}%
\end{align}

In order to obtain the full vertex corresponding to the conserved
(axial-)vector current it is necessary to add the term which contains the
(axial-)vector meson propagator. The addition of this term to the full
conserved vertex acquires the form%
\begin{equation}
\Delta\Gamma_{\mu}^{\left(  5\right)  a}(p,q,p^{\prime}=p+q)=T^{a}\left[
-\left(  g_{\mu\nu}-\frac{q^{\mu}q^{\nu}}{q^{2}}\right)  \gamma_{\nu}%
\frac{G_{J}B_{J}\left(  q^{2}\right)  }{1-G_{J}J_{J}^{T}\left(  q^{2}\right)
}f^{V}\left(  p\right)  f^{V}\left(  p^{\prime}\right)  \right]  \left(
\gamma_{5}\right)  , \label{GVdress}%
\end{equation}
where the factors $B_{J}\left(  q^{2}\right)  $ define the virtual transition
of the meson to the (axial-)vector current and are given above in (\ref{BV})
and (\ref{BA}).

\section{Vector and axial-vector mesons properties}

To fit the parameters of the extended model (\ref{Lambda's}) and (\ref{G's}),
we consider the main decay modes of (axial-) vector mesons. The quark-meson
couplings are fixed by the $\rho,$ $\omega$ and $a_{1}$ masses from the
condition (\ref{gM}) and equal to
\begin{equation}
g_{\rho q}=0.61,\quad g_{\omega q}=0.59,\quad g_{a_{1}q}=0.08. \label{gMq}%
\end{equation}
The decay $\rho\rightarrow\pi\pi$ is described by the amplitude
\begin{equation}
\left\langle \pi^{a}(p_{1})\pi^{b}(p_{2})\left\vert \rho^{c}(P)\right.
\right\rangle =i\varepsilon^{abc}g_{\rho\pi\pi}\left(  p_{2}-p_{1}\right)
_{\mu}\epsilon^{\mu}, \label{RhoPiPi}%
\end{equation}
where $p_{i}$ are momenta of pions, $\epsilon^{\mu}$ is the $\rho$-meson
polarization vector. With parameters given in Sect. 6 we obtain $g_{\rho\pi
\pi}=6.1$ and the decay width $\Gamma_{\rho\pi\pi}=154$ MeV which reasonably
well agrees with the experimental value $\left[  \Gamma_{\rho\pi\pi}\right]
_{\mathrm{exp}}=149.2\pm0.7$ MeV \cite{PDG}.

The decays of vector mesons to $e^{+}e^{-}$ pair and the transition of the
$a_{1}$ meson to the axial-vector current are described by the amplitudes%
\begin{equation}
\left\langle 0\left\vert J^{\mu a}\right\vert \rho_{s}^{b}\right\rangle
=-g_{\rho\gamma}\delta^{ab}\epsilon_{s}^{\mu},\quad\left\langle 0\left\vert
J^{\mu}\right\vert \omega_{s}\right\rangle =-g_{\omega\gamma}\delta
^{ab}\epsilon_{s}^{\mu},\quad\left\langle 0\left\vert J^{\mu5}\right\vert
a_{1s}^{b}\right\rangle =-g_{a_{1}}\delta^{ab}\epsilon_{s}^{\mu}.
\label{gMgamma}%
\end{equation}
We have obtained the values for the photon-vector meson couplings
$g_{\rho\gamma}=0.114$ GeV$^{2},\quad g_{\omega\gamma}=0.039$ GeV$^{2}$ and
the axial coupling $g_{a_{1}}=0.082$ GeV$^{2}$ which have to be compared with
the empirical values $g_{\rho\gamma}^{\exp}=0.1177\quad$GeV$^{2},$
$g_{\omega\gamma}^{\exp}=0.0359$ GeV$^{2}$ \cite{PDG}.

\section{Non-chiral corrections}

Here we give the expressions for nonchiral corrections. Hadronic spectral
functions for the processes $\rho\rightarrow\pi\pi\rightarrow\rho$ and
$a_{1}\rightarrow\sigma\pi\rightarrow a_{1}$ are given by general expression:%
\begin{equation}
\frac{1}{t}\frac{1}{\pi}\operatorname{Im}\Pi\left(  t\right)  =\frac{1}%
{3\cdot16\pi^{2}}\sqrt{\left(  1-\frac{m_{1}^{2}+m_{2}^{2}}{t}\right)
^{2}-2\frac{m_{1}^{2}+m_{2}^{2}}{t^{2}}}\left[  \left(  1-\frac{m_{1}%
^{2}+m_{2}^{2}}{t}\right)  ^{2}-\frac{4m_{1}^{2}m_{2}^{2}}{t^{2}}\right]
\Theta\left(  t-\left(  m_{1}+m_{2}\right)  ^{2}\right)  ,
\end{equation}
where $m_{1,2}$ are meson masses in the intermediate state. One has
$m_{1}=m_{2}=m_{\pi}$ for the first process and $m_{1}=m_{\pi},m_{2}%
=m_{\sigma}$ for the second one. The contribution of chiral loops to the
vector polarization function reads as%
\begin{equation}
\Pi_{\chi\mathrm{Loop}}^{V}\left(  Q^{2}\right)  =\frac{1}{48\pi^{2}}\left[
p\left(  \frac{Q^{2}}{4m_{\pi}^{2}}\right)  +p\left(  \frac{Q^{2}}{4m_{K}^{2}%
}\right)  \right]  ,
\end{equation}
where
\[
p\left(  z\right)  =-\left[  \frac{2}{3}\frac{3+4z}{z}+\left(  \frac{z+1}%
{z}\right)  ^{3/2}\left(  \operatorname{arctanh}\left[  \frac{1+2z}%
{2\sqrt{z\left(  z+1\right)  }}\right]  +i\frac{\pi}{2}\right)  \right]  .
\]

The leading order corrections to the low energy constants and to the
electromagnetic pion mass difference are correspondingly
\cite{HippeKl,KlevLemWil}%
\begin{equation}
\left[  I_{\mathrm{DMO}}\right]  _{\mathrm{\chi-loops}}=-\frac{1}{48\pi^{2}%
}\ln\frac{m_{\pi}^{2}}{m_{\sigma}^{2}},\qquad\left[  \left\langle r_{\pi}%
^{2}\right\rangle \right]  _{\mathrm{\chi-loops}}=\frac{3}{f_{\pi}^{2}}\left[
I_{\mathrm{DMO}}\right]  _{\mathrm{\chi-loops}},\qquad\left[  \Delta m_{\pi
}^{2}\right]  _{\mathrm{\chi-loops}}=-\frac{3\alpha}{4\pi}m_{\pi}^{2}\ln
\frac{m_{\pi}^{2}}{m_{\sigma}^{2}},\label{ChirCorr}%
\end{equation}
with chiral logarithm being approximately%
\begin{equation}
\ln\frac{m_{\pi}^{2}}{m_{\sigma}^{2}}\approx-3.
\end{equation}
Above estimate corresponds to the $\sigma-$pole position in $\pi\pi$ $S$-wave
isoscalar scattering around $600$ MeV.  In Table 2 we present separately
different terms contributing to the physical constants.

Table 2.%

\begin{tabular}
[c]{|c|c|c|c|c|c|}\hline
& quark loop & intermed. mesons & chiral loop & total & exp\\\hline
$\Delta m_{\pi}\left(  \mathrm{MeV}\right)  $ & $3.95$ & $0.13$ & $0.36$ &
$4.44$ & $4.43$\\\hline
$I_{\mathrm{DMO}}\cdot10^{3}$ & $16.72$ & $0.008$ & $6.16$ & $22.89$ &
$26.30$\\\hline
$\left\langle r_{\pi}^{2}\right\rangle \left(  \mathrm{fm}^{2}\right)  $ &
$0.334$ & $0.013$ & $0.087$ & $0.435$ & $0.439$\\\hline
\end{tabular}

\end{document}